\documentclass[sort&compress]{elsarticle}\usepackage{graphicx,color}
\usepackage{amsmath}
\usepackage{amssymb}

\newcommand \beq{\begin{eqnarray}}
\newcommand \eeq{\end{eqnarray}} 
\def\x{{\boldsymbol x}}
\def\a{{\boldsymbol a}}
\def\y{{\boldsymbol y}}

\def\r{{\boldsymbol r}}

\def\q{{\boldsymbol q}}
\def\s{{\boldsymbol s}}
\def\p{{\boldsymbol p}}
\def\y{{\boldsymbol y}}

\def\P{{\boldsymbol P}}
\def\X{{\boldsymbol X}}
\def\Y{{\boldsymbol Y}}

\def\bra#1{\langle#1\vert}
\def\ket#1{\vert#1\rangle}

\newcommand{\rmd}{{\rm d}}
\newcommand{\rme}{{\rm e}}

\newcommand{\del}{\partial}
\newcommand{\nn}{\nonumber\\ }

\begin{document}
\begin{frontmatter}
\title{ The approach to equilibrium of a quarkonium in a quark-gluon plasma}
\author[cea]{Jean-Paul Blaizot}

\address[cea]{Institut de  Physique Th\'eorique,  Universit\'e Paris Saclay, 
        CEA, CNRS, 
	F-91191 Gif-sur-Yvette, France} 

\author[Jy]{Miguel Angel Escobedo}
\address[Jy]{Department of Physics, P.O. Box 35, FI-40014 University of Jyv\"{a}skyl\"{a}, Finland}
\begin{abstract}
We derive equations of motion for the reduced density matrix of a heavy quarkonium in contact with a quark-gluon plasma in thermal equilibrium. These equations allow in particular a proper treatment of the regime when the temperature of the plasma is comparable to the binding energy of the quarkonium. These equations are used to study how the quarkonium approaches equilibrium with the plasma, and we discuss the corresponding entropy increase, or free energy decrease, depending on the temperature regime. The effect of collisions can be accounted for by the generalization of the imaginary potential introduced in previous studies, and from which collision rates are derived.  An important outcome of the present analysis is that this imaginary potential has a sizeable dependence  on the energy of the relevant transitions. 
 
\end{abstract}


\end{frontmatter}


\section{Introduction}

There is an ongoing major effort to measure the production of heavy quark bound states  in heavy ion experiments (for a recent review, see \cite{Scomparin:2017pno}). The goal of such measurements is to obtain information on the medium in which these heavy quark systems evolve.  However, to achieve such a goal,  we need to have good control of the dynamics of heavy quarks in a plasma, which is a difficult many-body problem.
Different physical effects could play a role in modifying the properties of heavy quark bound states in a quark-gluon plasma, the most prominent ones being the  
 screening of the binding forces and the collisions of the heavy quarks with the plasma constituents. The various models used in phenomenological analysis emphasize one aspect or the other, with of course many refinements in either direction. It is important however that all aspects of the dynamics be  treated on the same footing, within a coherent formalism.  Only then can one get confident that we understand the processes considered, and eventually extract from the data  the properties of the medium in which the bound state evolves.

 Important progress in this direction has occurred in the last few years. A major step forward was the recognition that the effect of the collisions could be incorporated in an imaginary potential \cite{Laine:2006ns,Beraudo:2007ky,Brambilla:2008cx,Escobedo:2008sy,Escobedo:2010tu}, somewhat analogous 
 to the optical potential used in nuclear physics. This imaginary potential can then be calculated, albeit not yet with the same degree of accuracy as the real potential that is responsible for binding and is screened in a plasma. Attempts to access it via lattice calculations can be found for instance in \cite{Rothkopf:2011db,Burnier:2015tda}.
 As for the real  potential, effective field theories have been used to constrain it in some particular regimes  \cite{Escobedo:2008sy,Brambilla:2008cx,Escobedo:2010tu}. It has also been realized that techniques borrowed from the theory of open  quantum systems  (see e.g. \cite{OQS}) could offer a new perspective on these issues. In particular, the imaginary potential appears naturally in the construction of the operators of the  Lindblad equation \cite{Akamatsu:2014qsa}. The stochastic potential used in some approaches   in connection with a Schr\"{o}dinger equation  (see e.g. \cite{Rothkopf:2013kya,Gossiaux:2016htk}) is also intimately related to this imaginary potential. As we shall see in this paper,  the imaginary potential also directly enters the calculations of the relevant transition rates. 

The present paper complements the study presented in Ref.~\cite{Blaizot:2017ypk}. There, a complete derivation of the equation of motion for the reduced density matrix has been given, under the assumption that the intrinsic dynamics of the heavy quarks is slow compared to that of the plasma. This assumption allowed us to reduce the equations of motion to equations of a  Langevin type. This assumption is strictly valid in the regime of high temperature, where the effect of binding forces are small and can be incorporated in the Langevin dynamics. The results that had been obtained along the same  lines in the abelian case in \cite{Blaizot:2015hya} suggest that it is in this case a reasonable approximation, even when bound states can form.   However, this is not so in QCD: when a quarkonium absorbs or emits a gluon its color state changes, from singlet to octet or vice versa. Since the potential between a quark and an antiquark is attractive in the singlet channel, and repulsive in the octet channel, the absorption of a gluon leads to a significant change in the effective heavy quark hamiltonian.  This conflicts with some of the assumptions underlying the  derivations presented in \cite{Blaizot:2017ypk}, which need therefore to be revisited. More broadly, we need to address more precisely the regime of moderate temperature where the binding energy is of the order of the temperature. 

We consider in this paper a simplified set up, a static quark-gluon plasma in thermal equilibrium, and study the time evolution of a single heavy quarkonium in  such a medium. The paper contains three main parts. In the next section, we derive equations of motion for the reduced density matrix for a quark-antiquark pair. These equations reproduce in some limit those obtained in \cite{Blaizot:2017ypk}, but they lend themselves to   more accurate approximations in the regime where the temperature is of the order of the binding energy of the quarkonium. These equations are simplified by integrating out the center of mass coordinates, leaving us with equations for the relative motion alone. The second part of the paper, which covers Sects.~\ref{sec:SQED} and \ref{sec:EFQCD}, presents a general discussion of how the quarkonium approaches  equilibrium with the quark-gluon plasma. We shall see that different treatments can be given depending on whether the temperature is large compared to the binding energy, or comparable to it. This will lead us to consider the variation with time of an (off-equilibrium) entropy and free energy. The third part of the paper, Sect.~\ref{sec:illustrations}, presents some numerical calculations illustrating the main features of the general equations in some simplified situations. Conclusions are summarized at the end.

\section{The evolution equation for the density matrix}

We consider a single heavy quark-antiquark pair immersed in a plasma of light quarks and gluons in thermal equilibrium at a temperature $T$ much smaller than the mass $M$ of the heavy quark. The condition $M\gg T$ ensures that we can treat the heavy quark and antiquark as  non-relativistic particles. Also, since the velocity of the heavy particles is small ($\lesssim \sqrt{T/M}$), we neglect their magnetic interactions (among themselves, and with the plasma constituents)\footnote{This means, in particular, that the processes of gluo-dissociation are left out of the present discussion. Including those would, however, amount to a  straighforward generalisation of the present formalism (see e.g. the footnote before Eq.~(\ref{eq.122})).}. We assume then that the whole system can be described by the following Hamiltonian
\begin{equation}\label{hamiltonian}
H=H_{\rm pl}+H_{Q}+H_1\,,
\end{equation}
where $H_{\rm pl}$ is the QCD Hamiltonian governing the dynamics of the plasma while $H_{Q}$ controls the dynamics of the heavy quark-antiquark pair in the absence of the plasma.  The hamiltonian $H_Q$ reads
\begin{equation}\label{HQhamiltonian}
H_{Q}=H_{\rm s,o}=-\frac{\Delta_{\boldsymbol r}}{M}-\frac{\Delta_{\boldsymbol R}}{4M}+V_{\rm s,o}({\boldsymbol r})\,,
\end{equation}
where  ${\boldsymbol r}$ and ${\boldsymbol R}$ denote respectively the relative and the center of mass coordinates of the heavy particles. The interaction potential $V_{\rm s,o}({\boldsymbol r})$ is a function of the relative coordinates,  and it depends also on the color configuration of the pair. Thus, as indicated in Eq.~(\ref{HQhamiltonian}), we shall often write $H_Q$ as either $H_{\rm s}$ or $H_{\rm o}$, depending on whether the quark-antiquark pair is in a color singlet ($H_{\rm s}$)  or a color octet ($H_{\rm o}$) configuration. In leading order non-relativistic limit, i.e., keeping only the color Coulomb interaction into account, we have
\begin{equation}\label{eq:Vso}
 V_{\rm s}({\boldsymbol r})=-\frac{C_F\alpha_s }{r},\qquad V_{\rm 0}({\boldsymbol r})=\frac{\alpha_s }{2N_cr},
\end{equation}
where $C_F=(N_c^2-1)/(2N_c)$, with $N_c=3$ the number of colors, and $\alpha_s$ is the strong coupling constant, $\alpha_s=g^2/(4\pi)$ with $g$ the gauge coupling. 

The last term in Eq.~(\ref{hamiltonian}) is the interaction between the plasma and the heavy quarks. It is of the form\footnote{Throughout this paper, we use the shorthand notation  $\int_\x\equiv \int \rmd^3 \x$ for the spatial integrals, and $\int_\p\equiv \int \frac{\rmd^3\p}{(2\pi)^3}$ for momentum integrals.}
\beq\label{H1}
H_1=-g\int_\x a_0^A(\x) n^A(\x),
\eeq
where $a_0^A$ denotes the (color) Coulomb field created by the plasma particles, while $n^A$ denotes the color charge density of the heavy particles, with $A$ a color index. For a quark-antiquark pair, the color charge density is given by
\beq\label{colordensity}
n^A(\x)=\delta(\x-\hat {\boldsymbol r})\, T^A\otimes \mathbb{I} -\mathbb{I}\otimes\delta(\x-\hat{\boldsymbol r}) \, \tilde T^A,
\eeq
where $\hat {\boldsymbol r}$ denotes the position operator\footnote{ We occasionally put a hat on operators whenever confusion may arise from not doint so.}, and  the two components of the tensor product refer respectively to the Hilbert space of the heavy quark (for the first component) and the heavy antiquark (for the second component).
In Eq.~(\ref{colordensity}), $T^A$ is a color matrix in the fundamental representation of SU(3) and describes the coupling between the heavy quark and the gluon. The coupling of the heavy antiquark and the gluon is described by $-\tilde T^A$, with $\tilde T^A$ the transpose of $T^A$.\\

\subsection{The reduced density matrix and its color structure}

Consider now the density matrix ${\cal D}$ of the whole system. We assume that initially, at time $t_0$, this density matrix factorizes
\beq
{\cal D}(t_0)={\cal D}_Q(t_0)\otimes{\cal D}_{\rm pl}(t_0),
\eeq
where the plasma density matrix ${\cal D}_{\rm pl}(t_0)$  is an equilibrium density matrix at temperature $T=1/\beta$:
\beq
{\cal D}_{\rm pl}(t_0)=\frac{\rme^{-\beta H}}{Z_{\rm pl}},\qquad Z_{\rm pl}={\rm Tr}\rme^{-\beta H_{\rm pl}}.
\eeq
The reduced density matrix, ${\cal D}_Q$, the objet that we are mostly concerned with,  is defined  by taking the trace over the plasma degrees of freedom
\begin{equation}
{\cal D}_{Q}(t)={\rm Tr}_{\rm pl} ({\cal D}(t)).
\end{equation}

The state of a heavy quark can be characterized by a position, a color, and a spin. We ignore here  the spin degree of freedom. Then the reduced density matrix ${\cal D}_Q$ has matrix elements of the form
\beq
\langle {\boldsymbol r}_1a, \bar{\boldsymbol r}_1 \bar a |{\cal D}_Q|{\boldsymbol r}_2 b,\bar{\boldsymbol r}_2 \bar b\rangle,
\eeq
where $a, b$  and $\bar a, \bar b$   are color indices in the fundamental representation and  its conjugate, respectively, while ${\boldsymbol r}_i$ and $\bar{\boldsymbol r}_i$  ($i=1,2$) denote respectively  the coordinates of the quark and the antiquark. 
Factorizing the color structure, one can write ${\cal D}_Q$ as follows (see \cite{Blaizot:2017ypk} for more details on the color structure of ${\cal D}_Q$). 
\beq
{\cal D}_{Q}(t)&=&\left(\frac{\delta_{a\bar{a}}\delta_{b\bar{b}}}{N_c}{D}_{\rm s}(t)+\frac{T^A_{a\bar{a}}T^A_{\bar{b}b}}{T_F}{D}_{\rm o}(t)\right)|a,\bar{a}\rangle\langle b,\bar{b}|\nn
&=& D_{\rm s}(t) \ket{\rm s}\bra{\rm s}+D_{\rm o}(t) \sum_C \ket{\rm o^C}\bra{\rm o^C}
\label{eq:rhodecom1}
\eeq
where ${D}_{\rm s }$ and ${D}_{\rm o}$ are matrices in the 2 particle space, with only coordinates as entries, e.g. the matrix elements of ${D}_{\rm s}$ are $\langle {\boldsymbol r}_1, \bar{\boldsymbol r}_1 |{D}_{\rm s}|{\boldsymbol r}_2 ,\bar{\boldsymbol r}_2 \rangle$. In the formula above, $T_F=1/2$ and the color matrices are normalized as ${\rm Tr}\,T^A T^B=\delta^{AB}/2$. The relation between the first and second lines of Eq.~(\ref{eq:rhodecom1}) follows from the   following formulae 
\beq
\bra{a\bar a}{\rm s}\rangle=\delta_{a\bar a}\frac{1}{\sqrt{N_c}},\qquad  \bra{a\bar a}{\rm o^C}\rangle=\sqrt{2} \,T^C_{a\bar a},
\eeq
where $\ket{\rm s}$ and $\ket{{\rm o}^C}$ denote respectively color singlet and octet (normalized) states, the index {\scriptsize\it C} in  ${\rm o}^C$ being a color index that distinguishes the various members of the octet. 
In the  limit where the mass of the heavy quark is infinite, the density matrix is diagonal in coordinate space, 
\beq\label{DlargeM}
\langle {\boldsymbol r}_1, \bar{\boldsymbol r}_1 |D_{\rm s}|{\boldsymbol r}_2 ,\bar{\boldsymbol r}_2 \rangle=\delta({\boldsymbol r}_1-{\boldsymbol r}_2)\delta(\bar{\boldsymbol r}_1-\bar{\boldsymbol r}_2)D_{\rm s}({\boldsymbol r}_1-\bar{\boldsymbol r}_1),
\eeq
and similarly for $D_{\rm o}$. In this limit the density matrix depends only on the relative coordinate ${\boldsymbol r}_1-\bar{\boldsymbol r}_1$, which follows from the fact that the plasma in equilibrium is invariant under translations. \\

\subsection{Approximate evolution equation for the reduced density matrix}

 The time evolution of the density matrix of the full system obeys the general equation of motion
 \beq
 i\frac{\rmd {\cal D}}{\rmd  t}=[H,{\cal D}].
 \eeq
 In order to treat the interaction between the plasma and the heavy particles by using perturbation theory, we move to the interaction representation with respect to the unperturbed hamiltonian $H_0=H_Q+H_{\rm pl}$ and define 
 \beq
 {\cal D}(t)=U_0(t,t_0) {\cal D}^I(t) U_0^\dagger(t,t_0),
 \eeq
 where ${\cal D}^I(t) $ satisfies the equation
 \beq\label{exacteqintrepres0}
 \frac{\rmd {\cal D}^I}{\rmd  t}&=&-i [H_1(t),{\cal D}^I(t)],\qquad  H_1(t)=U_0(t,t_0)^\dagger H_1 U_0(t,t_0).
 \eeq
 We can then rewrite the equation of motion (\ref{exacteqintrepres0}) as
  \beq\label{exacteqintrepres1}
 \frac{\rmd {\cal D}^I}{\rmd  t}=-i [H_1(t),{\cal D}^I(t_0)]-\int_{t_0}^t\rmd t' [H_1(t),[H_1(t'),{\cal D}^I(t')]].
 \eeq
 This exact equation is obtained by formally integrating Eq.~(\ref{exacteqintrepres0}) and inserting  the obtained solution  back into the equation. Perturbation theory at second order in $H_1$ is recovered by replacing ${\cal D}^I(t')$ by ${\cal D}^I(t_0)$ in the double commutator. 
 
 We can however improve on strict perturbation theory, with the help of two approximations. These are  consistent with strict second order perturbation theory but go beyond, in particular by performing partial resummations (analogous to those in Schwinger-Dyson equations). 
The first approximation consists in replacing in the double commutator in the  right hand side of Eq.~(\ref{exacteqintrepres1})  the density matrix by the factorized form
 \beq\label{factorizedD}
 {\cal D}^I(t)= {\cal D}_Q^I(t)\otimes  {\cal D}_{\rm pl}^I(t).
 \eeq
 This is consistent with second order perturbation theory since deviation from this form necessarily involves extra powers of $H_1$. The factorization  allows us to perform easily the average over the plasma degrees of freedom. We then obtain the following equation for the reduced density matrix of the heavy quarks
\beq\label{eqrhoAt0b2b}
\frac{\rmd {\cal D}^I_Q(t)}{\rmd t}
&=&-g^2\int_{t_0}^t \rmd t'\int_{\x\x'}\left( [n^A(t,\x), n^A(t',\x'){\cal D}^I_Q(t')]\Delta^>(t-t',\x-\x')\right.\nn
&&\qquad\qquad\quad \left.  +[{\cal D}^I_Q(t') n^A(t',\x'),n^A(t,\x)]\Delta^<(t-t',\x-\x')\right).\nn
\eeq
We have used the fact that the linear term vanishes in  a neutral plasma, and the sum over the color index $A$ is implicit. Finally, we have written the correlator of the $a_0$ fields as
\beq\label{correlators}
&&{\rm Tr}_{\rm pl}\left[a_0^A(t,{\bf x})a_0^B(t',{\bf y}){\cal D}_{\rm pl}\right]=\delta^{AB}\Delta^>(t-t',{\bf x}-{\bf y}),\nn
&&{\rm Tr}_{\rm pl}\left[a_0^B(t',{\bf y})a_0^A(t,{\bf x}){\cal D}_{\rm pl}\right]=\delta^{AB}\Delta^<(t-t',{\bf x}-{\bf y}).
\eeq

\vspace{-0cm}
\begin{figure}[!hbt]
\begin{center}
\includegraphics[width=1\textwidth]{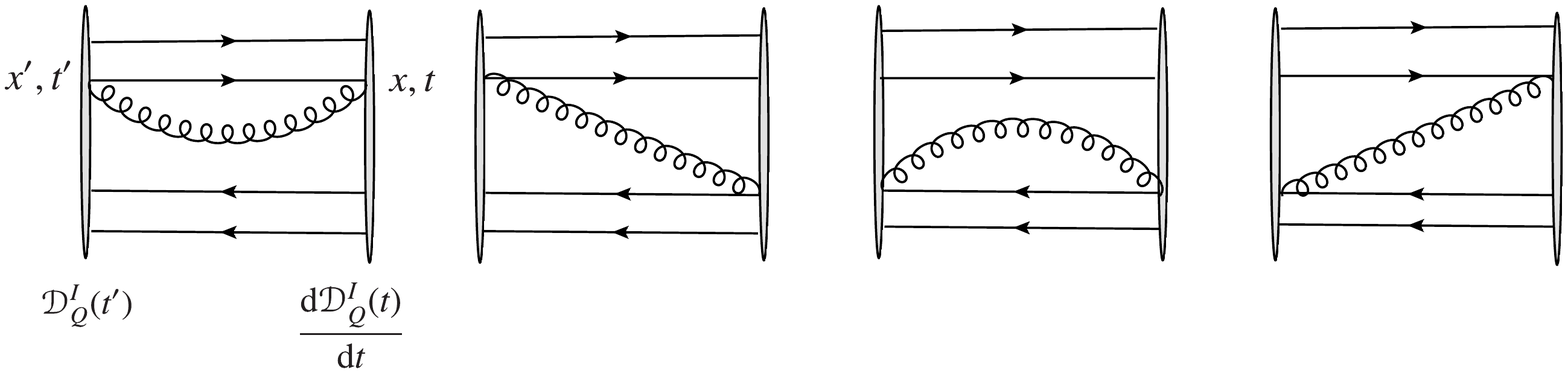}
\vspace{-5cm}
\caption{These four diagrams are in one-to-one correspondence with the four terms in  Eq.~(\ref{eqrhoAt0b2b}). The time flows as in a Schwinger-Keldysh contour, forward in the upper part, and backward in the lower part. The two upper lines represent the quark and the antiquark propagating from  $t'$ to $t$, while the lower lines represent the same particles  propagating from $ t$ to $t'$.  } \vspace{-0.25in}
\label{fig:Ws0}
\end{center}
\end{figure}

The equation (\ref{eqrhoAt0b2b}) can be given a simple diagrammatic interpretation, illustrated in Fig.~\ref{fig:Ws0}  (see \cite{Blaizot:2017ypk} for more details). The diagrams involve   single gluon exchanges, represented by the correlators (\ref{correlators}), with the gluon attached at points $(t',\x')$ and $(t,\x)$. Contributions where the two densities are on the same side of the density matrix in Eq.~(\ref{eqrhoAt0b2b}), like in $n^A(t,\x) n^A(t',\x'){\cal D}^I_Q(t')$, are associated to diagrams where the gluon joins lower or upper particle lines among themselves (the first and third diagrams in Fig.~\ref{fig:Ws0}). Contributions where a density is lying on each side of ${\cal D}$, as in  in $n^A(t,\x) {\cal D}^I_Q(t')n^A(t',\x')$, are represented by diagrams where the gluon joins upper and lower lines (the second and fourth diagram in Fig.~\ref{fig:Ws0}).

 The equation (\ref{eqrhoAt0b2b}) contains a non trivial memory integral. However a second approximation allows us to obtain a Markovian equation. Indeed, we note that  the difference  ${\cal D}^I_Q(t')- {\cal D}^I_Q(t)$ involves powers of the interaction. Thus, at the order at which we are working, we can  neglect this difference in the right hand side of Eq.~(\ref{eqrhoAt0b2b}), and simply substitute there ${\cal D}^I_Q(t')$ with ${\cal D}^I_Q(t)$. At this point, the equation still contains a non trivial time integral, but it is Markovian.  Moving back to the Schr\"odinger picture, one can write  this equation as 
 \beq\label{eqrhoAt0c2b2b2}
&&\!\!\!\!\!\!\!\!\!\! \frac{\rmd {\cal D}_Q}{\rmd t}+i[H_Q,{\cal D}_Q(t)]=\nn
&&-g^2 \int_{\x\x'}
 \int_{0}^{t-t_0} \rmd \tau \,[n^A_\x ,U_Q(\tau)n^A_{\x'}U_Q^\dagger(\tau) {\cal D}_Q(t)]  \, \Delta^>(\tau;\x-\x'))\nn
&&-g^2\int_{\x\x'}\int_{0}^{t-t_0} \rmd \tau \, [{\cal D}_Q(t)U_Q(\tau) n^A_{\x'}U_Q^\dagger(\tau), n^A_\x]  \,\Delta^<(\tau;\x-\x'),
\eeq
 where we have made a change of variable in the time integration, and set $t-t'=\tau$. 
This   Markovian  equation is the equation that was studied in Ref.~\cite{Blaizot:2017ypk}, and it will prove useful later on. In the rest of this section though, we shall use Eq.~(\ref{eqrhoAt0b2b}), which has a simpler diagrammatic interpretation. In fact, most of the derivations in this section are blind to this modification of the equation, as it will be clear. In the Schr\"odinger picture, Eq.~(\ref{eqrhoAt0b2b}) reads
\beq\label{eqrhoAt0c2b2b}
&&\!\!\!\!\!\!\!\!\!\! \frac{\rmd {\cal D}_Q}{\rmd t}+i[H_Q,{\cal D}_Q(t)]=\nn
&&-g^2 \int_{\x\x'}
 \int_{0}^{t-t_0} \rmd \tau \,[n^A_\x ,U_Q(\tau)n^A_{\x'}{\cal D}_Q(t-\tau)U_Q^\dagger(\tau)]  \, \Delta^>(\tau;\x-\x'))\nn
&&-g^2\int_{\x\x'}\int_{0}^{t-t_0} \rmd \tau \, [U_Q(\tau) {\cal D}_Q(t-\tau)n^A_{\x'}U_Q^\dagger(\tau), n^A_\x]  \,\Delta^<(\tau;\x-\x').
\eeq

\subsection{Equation of motion for $D_{\rm s}$ and $D_{\rm o}$}

  We shall write Eq.~(\ref{eqrhoAt0c2b2b}) as follows
\begin{equation}\label{mastereq}
\frac{\rmd}{\rmd t}{\cal D}_{Q}(t)=-i[H,{\cal D}_{Q}(t)]+\int_{0}^{t-t_0}\,\rmd \tau\,{\cal L}(\tau){\cal D}_Q(t-\tau),
\end{equation}
where ${\cal L} $ is to be understood as a linear operator acting on the density matrix. It corresponds typically to one gluon exchange processes (see Fig.~\ref{fig:Ws0}), with the gluon being emitted at time $t-\tau$ and absorbed at time $t$.  

Given the color structure of the density matrix (see Eq.~(\ref{eq:rhodecom1})), it is convenient to view ${\cal L}$ as a matrix in the 2-dimensional space spanned by the two components $D_{\rm s}$ and $D_{\rm o}$ of the density matrix.  Thus, we write 
\begin{eqnarray}\label{DsDoME}
\frac{\rmd D_{\rm s}}{\rmd t}=-i[H_{\rm s},D_{\rm s}]+\int_{0}^{t-t_0}\rmd\tau\left\{{\cal L}^{\rm ss}(\tau)D_{\rm s}(t-\tau)+{\cal L}^{\rm so}(\tau)D_{\rm o}(t-\tau)\right\},\nn
\frac{\rmd D_{\rm o}}{\rmd t}=-i[H_{\rm o},D_{\rm o}]+\int_{t_0}^t\,dt'\left\{{\cal L}^{\rm os}(\tau)D_{\rm s}(t-\tau)+{\cal L}^{\rm oo}(\tau)D_{\rm o}(t-\tau)\right\}. 
\end{eqnarray}
In order to perform the color algebra needed to obtain the explicit expressions of the operators ${\cal L}^{ij}$, we note that both the density matrix and the heavy quark hamiltonian are diagonal in the singlet-octet basis (a  property that we have already used in writing Eq.~(\ref{DsDoME})). Furthermore, we note that the density operator $n^A(\x)$ can connect singlet to octet states, and also various octet  states among themselves. Its matrix elements are given by (see e.g. \cite{Blaizot:2017ypk})
\beq\label{colmelnA}
&&\bra{{\rm s}}n^A_\x\ket{{\rm o}^C}=\frac{\delta^{AC}}{\sqrt{2N_c}} \, n(\x),\nn
&&\bra{{\rm o}^D} n^A_{\x}\ket{{\rm o}^C}=\frac{1}{2} d^{DAC} \, n(\x)+\frac{i}{2} f^{DAC}\,m(\x),
\eeq
where 
\beq\label{nxmx}
&&n(\x)\equiv\delta(\x-\hat{\boldsymbol r})\otimes\mathbb{I} -\mathbb{I}\otimes \delta(\x-\hat{\boldsymbol r}), \nn
&&m(\x)\equiv(\x-\hat{\boldsymbol r})\otimes\mathbb{I} +\mathbb{I}\otimes \delta(\x-\hat{\boldsymbol r}).
\eeq
The calculation then proceeds easily by inserting closure relations in the singlet-octet basis at appropriate places in Eq.~(\ref{eqrhoAt0c2b2b}), for instance
\beq
&&\bra{{\rm s}}n^A_\x n^A_{\x'}\ket{{\rm s}}=\sum_C\bra{{\rm s}}n^A_\x\ket{{\rm o}^C}\bra{{\rm o}^C} n^A_{\x'}\ket{{\rm s}},\nn
&&\bra{{\rm o}^C}n^A_\x n^A_{\x'}\ket{{\rm o}^C}=\bra{{\rm o}^C} n^A_{\x}\ket{{\rm s}}\bra{{\rm s}}n^A_{\x'}\ket{{\rm o}^C}+\sum_D\bra{{\rm o}^C}n^A_\x\ket{{\rm o}^D}\bra{{\rm o}^D} n^A_{\x'}\ket{{\rm o}C},\nn
\eeq
and using the  following formulas to complete the color algebra
\beq\label{colalgebra}
f^{ABC} f^{ABD}=N_c\delta^{CD}, \qquad d^{ABC}d^{ABD}=\frac{N_c^2-4}{N_c}\delta^{CD}, \qquad d^{ABC}\delta^{AB}=0.
\eeq

\begin{figure}[!hbt]
\begin{center}
\includegraphics[width=1\textwidth]{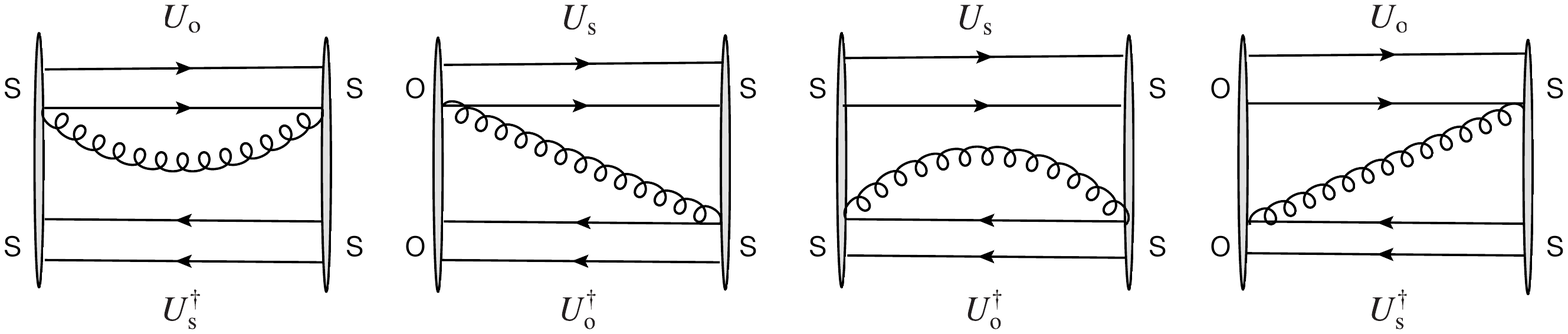}
\vspace{-5cm}
\caption{These four diagrams are in one-to-one correspondence with the four terms in Eqs.~(\ref{eq:evosin1},\ref{eq:evosin2}). In these diagrams  the evolution operators concern the heavy quark pair (not the gluon). Each of these diagrams represents four similar diagrams where the gluon is hooked in  four possible ways without changing the topological structure (these four contributions are summarized by the propagators $\Delta_-^{<}$ and $\Delta_-^{>}$, cf. Eq.~(\ref{deltabigminus})). The first and third diagrams represent ${\cal L}^{\rm ss}$, the second and fourth ${\cal L}^{\rm so}$.} \vspace{-0.25in}
\label{fig:Ws}
\end{center}
\end{figure}

It is then straightforward, by taking matrix elements of  Eq.~(\ref{eqrhoAt0c2b2b}) in singlet or octet states,  to obtain the expressions for the operators ${\cal L}^{ij}$ in Eqs.~(\ref{DsDoME}). Thus, by taking matrix elements in singlet states, we get (with $t'\equiv t-\tau$)
\beq
&&{\cal L}^{\rm ss}(\tau)D_{\rm s}(t')=
-g^2C_F\int_{\bf X,X'}\left\{\Delta_-^>({X},{X}')\;{\cal P}_{\X}\,U_{\rm o}(\tau)\,{\cal P}_{\X'}\,D_{\rm s}(t')\,U_{\rm s}^\dagger(\tau) \right.\nonumber\\
&&\qquad\qquad\qquad\qquad\qquad\left.+\Delta_-^<({X},{X}')\; U_{\rm s}(\tau)D_{\rm s}(t'){\cal P}_{\X'}U_{\rm o}^\dagger(\tau) {\cal P}_{\X}\right\},
\label{eq:evosin1}
\end{eqnarray}
\beq
&&{\cal L}^{\rm so}(\tau)D_{\rm o}(t')=g^2C_F\int_{\bf X,X'}\left\{\Delta_-^>({X},{X}')\; U_{\rm s}(\tau){\cal P}_{\X'}D_{\rm o}(t')U_{\rm o}^\dagger(\tau) {\cal P}_{\X} \right.\nonumber\\
&&\qquad\qquad\qquad\qquad\quad\left.+\Delta_-^<({X},{X}')\;{\cal P}_{\X}U_{\rm o}(\tau)D_{\rm o}(t'){\cal P}_{\X'}U_{\rm s}^\dagger(\tau) \right\}.
\label{eq:evosin2}
\end{eqnarray}
In these equations,  
\begin{equation}\label{deltabigminus}
\Delta^>_-({X},{X}')=\Delta^>(\tau,{\x}-{\x}')+\Delta^>(\tau,\bar{\x}-\bar{\x}')-\Delta^>(\tau,{\x}-\bar{\x}')-\Delta^>(\tau,\bar{\x}-{\x}'),
\end{equation}
and similarly for $\Delta^<_-({X},{X}')$. This expression represents the combination of propagators that naturally emerges when one adds the four possible ways to hook the gluon in diagrams with a given topology (i.e. in one of the diagrams of Fig.~\ref{fig:Ws}). The minus sign in the last two terms finds its origin in the minus sign present in $n(\x)$ in Eq.~(\ref{nxmx}) and affects the contributions where the gluon couples a quark and an antiquark. 
Furthermore, we have introduced the notation $X=(t,\X)$, where
$\X$ represents the set of coordinates of the quark-antiquark pair, i.e., $\X=\{{\x},\bar{\x}\}$, with coordinates without (with) bar giving the position of the quark  (antiquark). The integral $\int_\X$ in Eq.~(\ref{eq:evosin2}) runs over these coordinates, i.e., $\int_\X=\int\rmd^3\x\,\rmd^3\bar{\x}$.  Finally,  ${\cal P}_{\X}=|{\X}\rangle\langle{\X}|$ is a projector,  whose matrix elements between two localized states read
\begin{equation}
\bra{{\boldsymbol r}_1,\bar{\boldsymbol r}_1}{\cal P}_{\X}\ket{{\boldsymbol r}_2,\bar{\boldsymbol r}_2}= \delta({\boldsymbol r}_1-{\boldsymbol r}_2)\delta(\bar{\boldsymbol r}_1-\bar{\boldsymbol r}_2)\delta(\x-{\boldsymbol r}_1)\delta(\bar \x-\bar{\boldsymbol r}_1).
\end{equation}
\vspace{-0.5cm}

The evolution operators now depend on the color state of the propagating quark-antiquark pair. They are  $U_{\rm o}(\tau)=\rme^{-iH_{\rm o}\tau}$ for an octet state and $U_{\rm s}(\tau)=\rme^{-iH_{\rm s}\tau}$ for a singlet state. The operator $U(\tau)$ propagates the quark-antiquark pair forward in time, i.e., from $t'=t-\tau$ to $t$, while its hermitian conjugate, $U^\dagger(\tau)$ propagates the pair backward in time, from $t$ to $t-\tau$. The two operators are therefore attached  respectively to the upper and lower pairs of lines in diagrams such as those introduced in Fig.~\ref{fig:Ws0}. The structure of Eqs.~(\ref{eq:evosin1}, \ref{eq:evosin2}) may then be understood with the help of the diagrams  displayed in Fig.~\ref{fig:Ws}.

\vspace{-0cm}
\begin{figure}[!hbt]
\begin{center}
\includegraphics[width=1\textwidth]{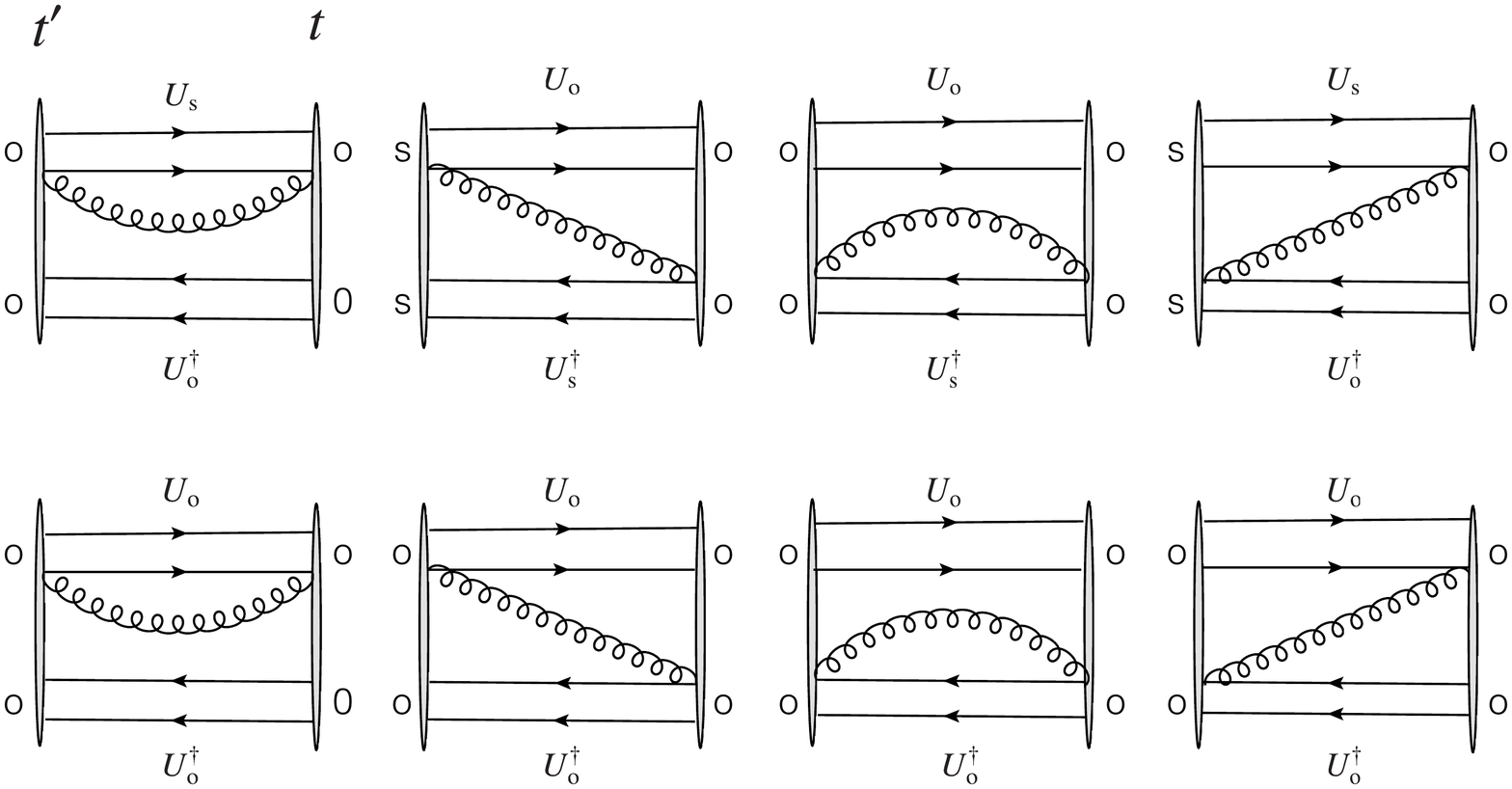}
\vspace{-2cm}
\caption{These four diagrams are in correspondence with the various terms in  Eqs.~(\ref{eq:evosin3} - \ref{eq:evosin4}). For instance the second and fourth diagrams in the first line correspond to ${\cal L}^{{\rm os}}$, the first and third  to ${\cal L}_1^{{\rm oo}}$, while the diagrams in the second line represent the various contributions to ${\cal L}_2^{{\rm oo}}$ and ${\cal L}_3^{{\rm oo}}$.} \vspace{-0.25in}
\label{fig:Wo}
\end{center}
\end{figure}

For the operators ${\cal L}^{{\rm o}j}$, we get
\beq\label{eq:evosin3}
&&{\cal L}^{\rm os}(\tau)D_{\rm s}(t')=\frac{g^2}{2N_c}\int_{\bf X,X'}\left\{\Delta_-^>({X},{X}')\; U_{\rm o}(\tau) {\cal P}_{\X'}D_{\rm s}(t')U_{\rm s}^\dagger(\tau) {\cal P}_{\X} \right.\nonumber\\
&&\qquad\qquad \qquad\qquad\qquad \left.+\Delta_-^<({X},{X}')\;{\cal P}_{\X}U_{\rm s}(\tau) D_{\rm s}(t'){\cal P}_{\X'}U_{\rm o}^\dagger(\tau) \right\},
\eeq
and, writing ${\cal L}^{\rm oo}={\cal L}^{\rm oo}_1+{\cal L}^{\rm oo}_2+{\cal L}^{\rm oo}_3$, 
\beq
&&{\cal L}^{\rm oo}_1(\tau)D_{\rm o}(t')=-\frac{g^2}{2N_c}\int_{\bf X,X'}\left\{\Delta_-^>({X},{X}')\;{\cal P}_{{\bf X}}U_{\rm s}(\tau) {\cal P}_{\X'}D_{\rm o}(t')U_{\rm o}^\dagger(\tau)  \right.\nonumber\\
&&\qquad\qquad \qquad\qquad \qquad\left.+\Delta_-^<({X},{X}')\;U_{\rm o}(\tau)  D_{\rm o}(t'){\cal P}_{\X'}U_{\rm s}^\dagger(\tau)  {\cal P}_{\X}\right\},
\end{eqnarray}
\beq\label{eq:evosin4a}
&&{\cal L}^{\rm oo}_2(\tau)D_{\rm o}(t')=-\frac{g^2(N_c^2-4)}{4N_c}\int_{\bf X,X'}\left\{ \Delta_-^>({X},{X}')\;[{\cal P}_{\X},U_{\rm o}(\tau) {\cal P}_{\X'}D_{\rm o}(t')U_{\rm o}^\dagger(\tau)]\right.\nn
&&\qquad\qquad\qquad\qquad\qquad\left.+\Delta_-^<({X},{X}')\;[U_{\rm o} (\tau)D_{\rm o}(t'){\cal P}_{\X'}U_{\rm o}^\dagger(\tau),{\cal P}_{\X}]\right\},
\end{eqnarray}
\beq\label{eq:evosin4}
&&{\cal L}^{\rm oo}_3(\tau)D_{\rm o}(t')=-\frac{g^2N_c}{4}\int_{\bf X,X'}\left\{\Delta_+^>({X},{X}')\;[{\cal P}_{\X},U_{\rm o}(\tau){\cal P}_{\X'}D_{\rm o}(t')U_{\rm o}^\dagger(\tau)]\right.\nn
&&\qquad\qquad\qquad\qquad\left.+\Delta_+^<({X},{X}')\;[U_{\rm o}(\tau)D_{\rm o}(t'){\cal P}_{\X'}U_{\rm o}^\dagger(\tau),{\cal P}_{\X}]\right\},
\end{eqnarray}
where 
\begin{equation}
\Delta_+^>({X},{X}')=\Delta^>(\tau,{\x}-{\x}')+\Delta^>(\tau,\bar{\x}-\bar{\x}')+\Delta^>(\tau,{\x}-\bar{\x}')+\Delta^>(\tau,\bar{\x}-{\x}'),
\end{equation}
and similarly for $\Delta_+^<({X},{X}')$. Note that there is no minus sign in $\Delta_+^>({X},{X}')$. This is because this contribution arises from products of factors $m(\x)$ in Eq.~(\ref{nxmx}). As for the color factors, their origins can be easily traced back to Eqs.~(\ref{colmelnA} - \ref{colalgebra}). \\

The equations of motion that we have obtained for $D_{\rm s}$ and $D_{\rm o}$ are similar to those derived in Ref.~\cite{Blaizot:2017ypk}, to within the small change discussed above (see after Eq.~(\ref{eqrhoAt0c2b2b2})), and the fact that in Ref.~\cite{Blaizot:2017ypk} the hamiltonian $H_0$ used in the interaction representation contains only the heavy quark and antiquark kinetic energy. At that point, in Ref.~\cite{Blaizot:2017ypk}  a further approximation was performed, that consists in expanding the evolution operators at short time, i.e., writing $U(\tau)\simeq 1-iH_0\tau$. Here we shall proceed differently in our treatment of the time integrals. The need to go beyond the approximation used in Ref.~\cite{Blaizot:2017ypk} is motivated in particular by the color changing transitions that take place in QCD: when a quark-antiquark pair in a singlet state absorbs a gluon, it turns into an octet state. This produces an immediate change in the effective hamiltonian of the pair, the force between the quark and the antiquark turning from attractive in the singlet state to repulsive in the octet state. It is then important to keep track of this change of the pair hamiltonian as the pair propagates during the lifetime of the exchange gluon. This is precisely what the various evolution operators do in the equations written above  (see also Figs.~\ref{fig:Ws} and \ref{fig:Wo}), and why we have included the leading order interaction between the quark and the antiquark in the hamiltonian $H_Q$.


\subsection{Equations for the relative motion}
At this stage, before doing any further approximation, we shall first simplify the equations that we have obtained by eliminating the center of mass coordinates. 
To that aim, we define a further reduced density matrix by taking a trace over the center of mass coordinates.  With ${\boldsymbol r}$ and $\bar {\boldsymbol r}$ denoting respectively the coordinates of the quark and the antiquark, we call ${\bf R}=({{\boldsymbol r}+\bar{\boldsymbol r}})/{2}$  the center of mass coordinate and $\s={\boldsymbol r}-\bar {\boldsymbol r}$ the relative coordinate. The reduced density matrix is defined from the matrix elements $\bra{{\boldsymbol r}_1\bar{\boldsymbol r}_1}D_{\rm s,o}\ket{{\boldsymbol r}_2\bar{\boldsymbol r}_2}$ which, with a slight abuse of notation, we write also as $\bra{{\boldsymbol R}_1,\s_1}D_{\rm s,o}\ket{{\boldsymbol R}_2,\s_2}$. We call $\tilde D_{\rm s,o}$ the reduced density matrix obtained after taking the trace over the center of mass coordinates. That is, 
\beq
\bra{\s_1}\tilde D_{\rm s,o}\ket{\s_2}\equiv\int_{\boldsymbol R} \bra{{\boldsymbol R},\s_1}D_{\rm s,o}\ket{{\boldsymbol R},\s_2}.
\eeq
The derivation of the equations of motion for the reduced density matrices $D_{\rm s,o}$ is presented  in Appendix~\ref{sec:cm}. It is assumed there that the  center of mass  velocity is small, typically $\lesssim \sqrt{T/M}$. We obtain then the following equations.   For the singlet, we get
\beq
&& {\cal L}^{\rm ss}(\tau)\tilde D_{\rm s}(t')=-g^2C_F\left\{\Delta^>(q)\,{\cal S}_{\q\cdot\hat\s} U_{\rm o}(\tau){\cal S}_{\q\cdot\hat\s} \tilde D_{\rm s}(t')U_{\rm s}(\tau)^\dagger\right.\nonumber\\
&&\qquad\qquad\qquad\left.+\Delta^<(q)\,U_{\rm s}(\tau) \tilde D_{\rm s}(t'){\cal S}_{\q\cdot\hat\s} U_{\rm o}^\dagger(\tau){\cal S}_{\q\cdot\hat\s}\right\},
\label{eq:evorhos1}
\end{eqnarray}
and
\beq
&&{\cal L}^{\rm so}(\tau)\tilde D_{\rm o}(t-\tau)=g^2C_F\left\{\Delta^>(q)\,U_{\rm s}(\tau){\cal S}_{\q\cdot\hat\s} \tilde D_{\rm o}(t')U_{\rm o}^\dagger(\tau) {\cal S}_{\q\cdot\hat\s}\right.\nonumber\\
&&\qquad\qquad\qquad\left.+\Delta^<(q)\,{\cal S}_{\q\cdot\hat\s} U_{\rm o}(\tau) \tilde D_{\rm o}(t'){\cal S}_{\q\cdot\hat\s} U_{\rm s}^\dagger(\tau)\right\}.
\label{eq:evorhos2}
\end{eqnarray}
In these equations (see Appendix~\ref{sec:cm})
\beq\label{calS}
{\cal S}_{\q\cdot \s}\equiv 2\sin(\q\cdot \hat\s/2),
\eeq
where $\hat\s$ is the operator measuring the relative coordinate. 

 For the octet, we get
\begin{eqnarray}
&& {\cal L}^{\rm os}(\tau)\tilde D_{\rm s}(t')=\frac{g^2}{2N_c}\left\{\Delta^>(q)\,U_{\rm o}(\tau){\cal S}_{\q\cdot\hat\s} \tilde D_{\rm s} U_{\rm s}^\dagger (\tau){\cal S}_{\q\cdot\hat\s}\right.\nn
&&\qquad\qquad\left.+\Delta^<(q)\,{\cal S}_{\q\cdot\hat\s} U_{\rm s}(\tau)\tilde D_{\rm s}(t'){\cal S}_{\q\cdot\hat\s} U_{\rm o}^\dagger(\tau) \right\},\label{eq:evorhoo0}
\end{eqnarray}
\begin{eqnarray}
&&{\cal L}^{\rm oo}_1(\tau)\tilde D_{\rm o}(t')=-\frac{g^2}{2N_c}\left\{\Delta^>(q)\,{\cal S}_{\q\cdot\hat\s} U_{\rm s}(\tau){\cal S}_{\q\cdot\hat\s} \tilde D_{\rm o}(t')U_{\rm o}^\dagger(\tau)\right.\nn
&&\qquad\qquad\qquad\left.+\Delta^<(q)\,U_{\rm o}(\tau)\tilde D_{\rm o}(t'){\cal S}_{\q\cdot\hat\s} U_{\rm s}^\dagger(\tau){\cal S}_{\q\cdot\hat\s} \right\},
\label{eq:evorhoo1}
\end{eqnarray}
\begin{eqnarray}
&& {\cal L}^{\rm oo}_2(\tau)\tilde D_{\rm o}(t')=-\frac{g^2(N_c^2-4)}{4N_c}\left\{\Delta^>(q)[{\cal S}_{\q\cdot\hat\s}\,,U_{\rm o}(\tau){\cal S}_{\q\cdot\hat\s} \tilde D_{\rm o}(t')U_{\rm o}^\dagger(\tau)]\right.\nn
&&\qquad\qquad\qquad\qquad\qquad\left.+\Delta^<(q)[U_{\rm o}(\tau)\tilde D_{\rm o}(t'){\cal S}_{\q\cdot\hat\s}  U_{\rm o}^\dagger(\tau)\,,{\cal S}_{\q\cdot\hat\s}]\right\},
\label{eq:evorhoo}
\end{eqnarray}
\begin{eqnarray}
&& {\cal L}^{\rm oo}_3(\tau)\tilde D_{\rm o}(t')=-\frac{g^2 N_c}{4}\left\{\Delta^>(q)[{\cal C}_{\q\cdot\hat\s} \,,U_{\rm o}(\tau){\cal C}_{\q\cdot\hat\s} \tilde D_{\rm o}(t')U_{\rm o}^\dagger(\tau)]\right.\nn
&&\qquad\qquad\qquad\qquad\left.+\Delta^<(q)[U_{\rm o}(\tau) \tilde D_{\rm o}(t'){\cal C}_{\q\cdot\hat\s} U_{\rm o}^\dagger(\tau)\,,{\cal C}_{\q\cdot\hat\s} ]\right\},
\label{eq:evorhoo}
\end{eqnarray}
where we have set ${\cal C}_{\q\cdot\hat\s} \equiv 2\cos\left({\q}\cdot{\hat\s}/{2}\right)$. 
\\

The  equations above give a fairly complete account of the relative motion  of a heavy quark-antiquark pair in a static quark-gluon plasma in thermal equilibrium. They are however difficult to solve in full generality.
In order to get some familiarity with their physical content, we consider, in the next two sections the  general question of how they describe the approach to thermal equilibrium, in two distinct regimes: The first regime is that of high temperature, controlled by the increase of the entropy. In the second regime, where   the temperature is of the order of the binding energy, entropy effects compete with binding forces; there,  the relevant quantity to look at is a non equilibrium free energy which will be seen to decrease monotonously as the equilibrium is approached.   We analyze first  the Abelian case, and  in the following section we consider QCD. 

From now on, in order to alleviate the notation, we omit the tilde in $\tilde D_{\rm s,o}$ since we shall be dealing only with the reduced density matrix of the relative motion. 

\section{Entropy and free energy in the abelian case}
\label{sec:SQED}



 A derivation completely analoguous to the one done in the previous section gives, for the case of an abelian plasma,
\beq\label{mainQED}
&&\frac{\rmd {\cal D}}{\rmd t}=-i[H_Q,{\cal D}]+\int_{0}^{t-t_0}\rmd\tau \int_q e^{-iq_0\tau } {\cal L}(\tau) {\cal D}(t-\tau),
\eeq
with
\beq
&&{\cal L}(\tau) {\cal D}(t-\tau)=-g^2\left\{\Delta^>(q)[{\cal S}_{\q\cdot\hat{\boldsymbol r}},U(\tau){\cal S}_{\q\cdot\hat{\boldsymbol r}}{\cal D}(t-\tau) U^\dagger(\tau)]\right.\nn
&&\qquad\qquad\qquad\qquad \qquad\left.+\Delta^<(q)[U(\tau){\cal D}(t-\tau){\cal S}_{\q\cdot\hat{\boldsymbol r}} U^\dagger(\tau),{\cal S}_{\q\cdot\hat{\boldsymbol r}}]\right\}\,.
\label{eq:Abevo}
\end{eqnarray}
This equation is identical to the equation for $D_{\rm s}$ (Eq.~(\ref{DsDoME})) in which we replace $C_F$ by unity, set $D_{\rm o}=D_{\rm s}={\cal D}$, and use the expressions (\ref{eq:evorhos1}) and (\ref{eq:evorhos2}). Recall that ${\cal D}$ is the reduced density matrix of the quark-antiquark pair, after taking the trace over the center of mass coordinates. The operator ${\cal S}_{\q\cdot\hat{\boldsymbol r}}$ is an operator in the space of the relative coordinates (cf. Eq.~(\ref{calS})), with $\hat {\boldsymbol r}$ denoting the relative coordinate operator. Finally, in $H_Q$, the potential is the ordinary Coulomb potential, $V({\boldsymbol r})=-\alpha/r$ (analogous to $V_{\rm s}$ in Eq.~(\ref{eq:Vso}) from which it is deduced via the substitutions  $C_F\to 1$ and $\alpha_s\to \alpha$, with $\alpha$ the fine-structure  constant).

Our goal now is to perform, approximately, the integration over the time $\tau$, in order to simplify the right hand side of the equation of motion.  The strategies to do so differ, depending on whether we are in the high temperature limit of not, that is whether binding energy effects play an important role or not.

\subsection{The entropy increase at high temperature}\label{sec:QEDhighT}

In this subsection we focus on the high temperature regime, i.e., the regime where the temperature is much higher than the binding energy. In this regime, one can ignore the binding energy, and the approach to equilibrium is dominated by entropy effects. We shall indeed show that the equations of motion predict a monotonous increase of the entropy. These equations of motion lead naturally to quantum Brownian motion, which was studied more extensively in \cite{Blaizot:2017ypk}.  In fact  the equation for the reduced density matrix takes the form of a Lindblad equation, where the effect of the collisions are accounted for by an imaginary potential.

The temperature enters the equations through the gluon propagator, and limits the range of the $\tau$-integration to $\tau\lesssim 1/m_D$, with $m_D\lesssim T$.\footnote{In strick weak coupling, the Debye mass is of order $gT\ll T$. However, it is convenient to relax this condition and allow for values of $m_D$ as large as the temperature. In fact, in the numerical calculations presented later, we have typically $m_D\simeq 2T$.} The evolution operators $U(\tau)$  contribute phase factors of the form ${\rm e}^{-i\tau H}$, where    $H$ is the sum of a kinetic and a potential energy $V$. Collisions change the kinetic energy by an amount $q^2/M +Pq/M$, where $q\sim m_D$ is the typical momentum of a gluon exchanged in a soft collision, and $P\sim \alpha M$ the typical momentum of the heavy quark or antiquark in its relative motion in a Coulomb bound state.  The contribution to the phase factor coming from the change of the kinetic energy  is then of order $m_D/M\ll 1$ for the first term, and of order $\alpha$ for the second. In either case, these are small contributions that can be safely neglected. As for the binding force acting on the heavy quark, this can be estimated  as follows: for a Coulomb bound state,  we have parametrically, $\alpha/r\sim \alpha p\sim M\alpha^2$. Thus, when $M\alpha^2\lesssim m_D$, which can be satisfied at sufficiently high temperature, the potential energy $V$, and the binding energy, are small compared to the Debye mass,  and one can safely set $U(\tau)\approx 1$ in Eq.~(\ref{eq:Abevo}).\footnote{Note that this is a much stronger approximation than used in Ref.~\cite{Blaizot:2017ypk} where a linear expansion in $\tau$ was used.}  Along the same line, we assume that during the time interval between $t-\tau$ and $t$, the density matrix does not vary significantly so that we can replace ${\cal D}(t-\tau)$ by ${\cal D}(t)$  in the right hand side of Eq.~(\ref{eq:Abevo}) (alternatively, we could use the form (\ref{eqrhoAt0c2b2b2}) of the evolution equation for ${\cal D}$).  With these approximations the evolution equation greatly simplifies, and  reads
\beq\label{eqrhoAt0c2b2newQED3}
&&\frac{\rmd {\cal D}}{\rmd t}+i[H_Q,{\cal D}(t)]=\nn
&&\qquad-g^2 \int_{0}^{t-t_0} \rmd \tau \int_q\rme^{-iq_0\tau} \left(  \Delta^>(q)[{\cal S}_{\q\cdot\hat{\boldsymbol r}},{\cal S}_{\q\cdot\hat{\boldsymbol r}}  {\cal D}(t)]+\Delta^<(q)[{\cal D}(t) {\cal S}_{\q\cdot\hat{\boldsymbol r}}  ,{\cal S}_{\q\cdot\hat{\boldsymbol r}}] \right).\nn
\eeq
The integrand in the right hand side can be written as follows
\beq
&&\frac{1}{2}\left[{\cal S}_{\q\cdot\hat{\boldsymbol r}}^2,{\cal D}(t) \right]\,\left( \Delta^>(q)-\Delta^<(q)   \right)\nn
&&+\left(\frac{1}{2}\left\{ {\cal S}_{\q\cdot\hat{\boldsymbol r}}^2,{\cal D}(t) \right\}-{\cal S}_{\q\cdot\hat{\boldsymbol r}}{\cal D}(t){\cal S}_{\q\cdot\hat{\boldsymbol r}}\right)\left( \Delta^>(q)+\Delta^<(q)\right).
\eeq
The contributions of the first and the second lines are qualitatively different: as we shall see shortly, the first line yields a correction to the real part of the potential, and corresponds to hamiltonian evolution,  while the second line involves the imaginary part of the potential, and accounts in particular for dissipation. To see that, we perform the integration over $q_0$, and then the integration over $\tau$ in Eq.~(\ref{eqrhoAt0c2b2newQED3}). 
In the long time limit, $t-t_0\gg 1/m_D$, we can let the upper limit of the $\tau$-integration go to infinity. The $\tau$-integrations yield then (see \cite{Blaizot:2017ypk})
\beq
&&\int_0^\infty \rmd \tau\, g^2\Delta^>(\tau;\q)=\frac{i}{2}\delta V(\q)-\frac{1}{2} W(\q)\\
&&\int_0^\infty \rmd \tau\, g^2\Delta^<(\tau;\q)=-\frac{i}{2}\delta V(\q)-\frac{1}{2} W(\q).
\eeq
The correction $\delta V$ to the real part of the potential provides a contribution that adds up to the hamiltonian in the left hand side ($H_Q\to H_Q'$). This reads
\beq
-2i \int_\q \delta V(\q) \left[  {\cal S}_{\q\cdot{\boldsymbol r}}{\cal S}_{\q\cdot{\boldsymbol r}},{\cal D}(t) \right]&=&i\left[ \int_\q \delta V(\q) \left(\rme^{i\q\cdot{\boldsymbol r}}-1\right),{\cal D}\right]\nn
&=& i\left[\delta V({\boldsymbol r})-\delta V(0),{\cal D}\right],\eeq
where we have used $\delta V(-\q)=\delta V(\q)$.
This is the screening correction to the real part of the potential. In the Hard Thermal Loop (HTL) approximation, this is given by (see e.g. \cite{Beraudo:2007ky})
\beq
\delta V(r) -\delta V(0) = \frac{\alpha_s}{r}\left(  \rme^{-m_D r }-1 \right)+\alpha_s m_D,
\eeq
so that the total potential in $H_Q'$ is the screened potential $V=(\alpha_s/r)\exp(-m_D r)$ (to within an irrelevant constant term).

The remaining terms can be rearranged as follows
\beq\label{eqrhoAt0c2b2newQED4}
\frac{\rmd {\cal D}}{\rmd t}+i[H'_Q,{\cal D}(t)]=-  \int_\q  W(\q)\left(  {\cal S}_{\q\cdot\hat{\boldsymbol r}}  {\cal D}(t) {\cal S}_{\q\cdot\hat{\boldsymbol r}}-\frac{1}{2}\{{\cal D}(t), {\cal S}_{\q\cdot\hat{\boldsymbol r}}^2   \}\right).
\eeq
This equation has the structure of a Linblad equation \cite{Lindblad:1975ef}. It can be written as
\begin{equation}\label{eq:Linblad0}
\frac{\rmd {\cal D}}{\rmd t}=-i[H'_Q,{\cal D}]+\int_\q \left(L_{\bf q}{\cal D}(t)L^\dagger_{\bf q}-\frac{1}{2}\left\{{\cal D}(t),L^\dagger_{\bf q}L_{\bf q}\right\}\right),
\end{equation}
where the Lindblad operators  $L_{\bf q}$ take the form ($g^2\Delta^>(0,{\bf q})=-W(\q)\ge 0$)
\begin{equation}
L_{\bf q}=g\sqrt{\Delta^>(0,{\bf q})}\,{\cal S}_{\q\cdot{\boldsymbol r}} =L_\q^\dagger.
\label{eq:defCq}
\end{equation}

At this point,  we could rely on the theorem derived in \cite{Benatti:1987dz} for the Lindbald equation, in order to show that the entropy is a monotonically increasing function of time. This theorem requires that  $\int\frac{\,d^3q}{(2\pi)^3}\left(L^\dagger_{\bf q}L_{\bf q}-L_{\bf q}L^\dagger_{\bf q}\right)=0$, which clearly holds in our case since $L^\dagger_{\bf q}=L_{\bf q}$. However in order to highlight the difference between the present high temperature regime,  and the low temperature regime to be discussed in the next subsection, we shall proceed with an explicit and elementary derivation. 

From the  definition of the von Neumann entropy,
\begin{equation}\label{Sdefinition}
S(t)=-\rm{Tr}({\cal D}(t)\log{\cal D}(t))\,.
\end{equation}
and using the fact that  ${\rm Tr}{\cal D}(t))$ is independent of time, a property that can be verified explicitly on Eq.~(\ref{eq:Linblad0}), one easily obtains
\beq
\frac{\rmd S}{\rmd t}&=&-{\rm Tr}\left(\frac{\rmd {\cal D}}{\rmd t}\log{\cal D}\right)\nn
&=&-\int_\q {\rm Tr}\left[\left(L_{\bf q}\,{\cal D}(t)\,L_{\bf q}-\frac{1}{2}\left\{{\cal D}(t),L_{\bf q}L_{\bf q}\right\}\right)\log{\cal D}(t)\right].
\eeq
At this point, it is convenient to  use a representation   in which ${\cal D}(t)$  is diagonal, viz.
\begin{equation}
{\cal D}(t)=\sum_n p_n(t)|n\rangle\langle n|\,,
\end{equation}
with $p_n(t)\ge 0$, and  the states $n$ are function of time.
We then obtain
\begin{eqnarray}
\frac{\rmd S}{\rmd t}&=&-\int_\q\left(\sum_{n,m}p_n(t)\log p_m(t)|\langle n|L_{\bf q}|m\rangle|^2-\sum_n p_n(t)\log p_n(t)\langle n| L_{\bf q}L_{\bf q}|n\rangle\right)\nonumber\\
&=&\int_\q\sum_{n,m}|\langle n|L_{\bf q}|m\rangle|^2p_n(t)\log\left(\frac{p_n(t)}{p_m(t)}\right)\nonumber\\
&=&\frac{1}{2}\int_\q\sum_{n,m}|\langle n|L_{\bf q}|m\rangle|^2[p_n(t)-p_m(t)]\log\left(\frac{p_n(t)}{p_m(t)}\right).
\label{eq:proofSi}
\end{eqnarray}
This expression is manifestly positive, which implies that the entropy (\ref{Sdefinition}) indeed increases with time. As already emphasized, this proof is less general than the use of the theorem in \cite
{Benatti:1987dz} (it relies in particular on the property $L_q^\dagger=L_q$). \\

It is interesting to relate the rate of entropy increase to our function $W({\boldsymbol r})$, or equivalently $\Gamma({\boldsymbol r})$. 
 A crude estimate can be obtained as follows. First we note that
\begin{equation}
S(t)=-\sum_n p_n(t)\log(p_n(t))\,,
\end{equation}
therefore the combination $(p_n(t)-p_m(t))\log\left(\frac{p_n(t)}{p_m(t)}\right)$ that appears in Eq. (\ref{eq:proofSi}) is a priori of the same size as $S$. Also,  having in mind the approach to equilibrium of a system initially in a bound state, we may write (cf. Eq.~(\ref{eq:defCq}))
\beq
\langle n|L_{\bf q}|m\rangle&\approx &2g\sqrt{\Delta^>(0,{\bf q})}\sin\left(\frac{{\bf q}{\bf a_0}}{2}\right)A_{nm},
\eeq
where ${\a_0}$ is a vector whose modulus coincides with the Bohr radius of the bound state, and $A_{nm}$ are constants of order unity.  We can then  estimate the order of magnitude of the change of entropy as follows
\beq
\frac{1}{S}\frac{\rmd S}{\rmd t}&\approx& {g^2}\int_\q\Delta^>(0,{\bf q})\left({\cal S}_{\q\cdot\a_0} \right)^2\nn
&=& 2\int_\q W(\q) \left( \rme^{i\q\cdot\a_0}-1 \right)\nn
&=&2\Gamma(a_0),
\label{eq:Sch}
\eeq
where we have used $W(\q)=-g^2\Delta^>(0,\q)$, and $\Gamma({\boldsymbol r})=W({\boldsymbol r})-W(0)$, with $W({\boldsymbol r})$ the imaginary part of the potential. This estimate relates the rate of entropy increase to the imaginary part of the potential, that is to a typical collision rate, at a scale determined by the size of the bound state.

\subsection{Free energy minimization in the Abelian limit}
\label{sec:EQED}

Now we look at the regime in which the temperature and the binding energies are of the same order of magnitude. It is no longer legitimate to approximate the evolution operators by unit operators, as we did in the previous subsection.    In this case, the effects of entropy and binding compete. Then, a relevant object to look at is the non equilibrium generalization of the free energy.  To analyze the time dependence of this quantity one is naturally led to expand on a complete set of eigenstates of the hamiltonian. The equations of motion lead in this case to rate equations, with rates that identify with those obtained from Fermi golden's rule. One can still express the effect of the collisions, at least partially, through the imaginary part of a potential, provided one takes into account  the energy dependence of this potential.

We start again from Eq.~(\ref{mainQED}), with the right hand side written as in Eq.~(\ref{eqrhoAt0c2b2b2}), that is,
\beq
&&{\cal L}(\tau) {\cal D}(t-\tau)=-g^2\left\{\Delta^>(q)[{\cal S}_{\q\cdot\hat{\boldsymbol r}},U(\tau){\cal S}_{\q\cdot\hat{\boldsymbol r}}U^\dagger(\tau){\cal D}(t) ]\right.\nn
&&\qquad\qquad\qquad\qquad \qquad\left.+\Delta^<(q)[{\cal D}(t)U(\tau){\cal S}_{\q\cdot\hat{\boldsymbol r}} U^\dagger(\tau),{\cal S}_{\q\cdot\hat{\boldsymbol r}}]\right\}\,.
\label{eq:Abevo2}
\end{eqnarray}
In order to handle more easily the evolution operators $U(\tau)=\rme^{-iH_Q\tau}$ and $U^\dagger(\tau)$, we introduce at appropriate places projectors on eigenstates of $H_Q$, ${\cal P}_n=\ket{n}\bra{n}$,  and assume for simplicity absence of   degeneracy.  Note that, in contrast to the previous subsection, the states $n$ are now independent of time. We get
\beq
{\cal L}(\tau) {\cal D}(t-\tau)&=&-g^2\sum_{n,k}\left\{\Delta^>(q)\,\rme^{-i(E_n-E_k)\tau}[{\cal S}_{\q\cdot\hat{\boldsymbol r}},{\cal P}_n{\cal S}_{\q\cdot\hat{\boldsymbol r}}{\cal P}_k{\cal D}(t) ]\right.\nn
&& \qquad\left.+\Delta^<(q)\,\rme^{-i(E_n-E_k)\tau}[{\cal D}(t){\cal P}_n{\cal S}_{\q\cdot\hat{\boldsymbol r}}{\cal P}_k,{\cal S}_{\q\cdot\hat{\boldsymbol r}}]\right\}\,.
\label{eq:Abevo3}
\end{eqnarray}
The integration over $\tau $ can then be performed, using
\beq
\int_0^\infty\rmd\tau\,\rme^{-iq_0\tau}\rme^{-i(E_n-E_k)\tau}\rme^{-\epsilon\tau}=\frac{i}{E_k-E_n-q_0+i\epsilon}
\eeq
with $\epsilon\to 0^+$.
We get
\begin{eqnarray}
&&\frac{\rmd{\cal D}}{dt}+i[H,{\cal D}]=\nn
&&-i g^2\int_q\sum_{k,n}\left(\frac{\Delta^>(q)}{E_k-E_n-q_0+i\epsilon}[{\cal S}_{\q\cdot\hat{\boldsymbol r}},{\cal P}_n{\cal S}_{\q\cdot\hat{\boldsymbol r}}{\cal P}_k{\cal D}(t)]\right.\nonumber\\
&&\qquad\qquad\quad\left.+\frac{\Delta^<(q)}{E_k-E_n-q_0+i\epsilon}[{\cal D}(t){\cal P}_n{\cal S}_{\q\cdot\hat{\boldsymbol r}}{\cal P}_k,{\cal S}_{\q\cdot\hat{\boldsymbol r}}]\right).
\label{eq:evolt2}
\end{eqnarray}
In order to perform the integration over $q_0$ we note that
\beq
\int_{q_0}\frac{\Delta^>(q)}{E_k-E_n-q_0+i\epsilon}=-\frac{i}{2}\Delta^>(E_{kn},\q)-{\rm P}\int_{q_0}\frac{\Delta^>(q_0,\q)}{q_0-E_{kn}},
\eeq
where we have set $E_{kn}\equiv E_k-E_n$, and the symbol P in front of the integral denotes the principal value. Similarly, 
\beq
\int_{q_0}\frac{\Delta^<(q)}{E_k-E_n-q_0+i\epsilon}&=&\int_{q_0}\frac{\Delta^>(q)}{E_k-E_n+q_0+i\epsilon}\nn
&=&-\frac{i}{2}\Delta^>(E_{nk},\q)+{\rm P}\int_{q_0}\frac{\Delta^>(q_0,\q)}{q_0+E_{kn}}.
\eeq
We have used $\Delta^<(q_0,\q)=\Delta^>(-q_0,-\q)=\Delta^>(-q_0,\q)$, where the last equality follows from the rotational invariance of the plasma. We can then rewrite Eq.~(\ref{eq:evolt2}) as
\begin{eqnarray}
&&\frac{\rmd{\cal D}}{dt}+i[H_Q,{\cal D}]=\nn
&&-\frac{g^2}{2}\int_\q\sum_{k,n} \left(\Delta^>(E_{kn},\q)  [{\cal S}_{\q\cdot\hat{\boldsymbol r}},{\cal P}_n{\cal S}_{\q\cdot\hat{\boldsymbol r}}{\cal P}_k{\cal D}(t)]+\Delta^>(E_{nk},\q)[{\cal D}(t){\cal P}_n{\cal S}_{\q\cdot\hat{\boldsymbol r}}{\cal P}_k,{\cal S}_{\q\cdot\hat{\boldsymbol r}}]\right)\nonumber\\
&&+i g^2\int_\q\sum_{k,n}{\rm P}\int_{q_0}\frac{\Delta^>(q_0,\q)}{q_0-E_{kn}} [{\cal S}_{\q\cdot\hat{\boldsymbol r}},{\cal P}_n{\cal S}_{\q\cdot\hat{\boldsymbol r}}{\cal P}_k{\cal D}(t)]\nn
&&-i g^2\int_\q\sum_{k,n}{\rm P}\int_{q_0}\frac{\Delta^>(q_0,\q)}{q_0+E_{kn}} [{\cal D}(t){\cal P}_n{\cal S}_{\q\cdot\hat{\boldsymbol r}}{\cal P}_k,{\cal S}_{\q\cdot\hat{\boldsymbol r}}].\label{eq:evolt3}
\end{eqnarray}

In the case where the typical transitions involve energy differences that are small compared to the Debye mass, which controls the decay of $\Delta(q_0)$ with $q_0$, we can ignore the energies $E_{kn}$, and perform freely the sums over $n$ and $k$, which eliminates the projectors. Using the identities
\beq
{\rm P}\int_{q_0}\frac{g^2\Delta^>(q_0,\q)}{q_0} =-\frac{1}{2}\delta V(\q),\qquad g^2\Delta^>(0,\q)=-W(\q),
\eeq
one then easily recovers the result of the previous section, i.e., Eq.~(\ref{eqrhoAt0c2b2newQED4}).

 We return now to Eq.~(\ref{eq:evolt3}). In order to minimize the effects of the principal parts and focus on the dissipative part of the equation, we  use the eigenstates of $H'_Q$ instead of $H_Q$, and accordingly subtract the corresponding contribution of $\delta V$ in the right hand side of the equation. We get
\begin{eqnarray}
&&\frac{\rmd{\cal D}}{dt}+i[H'_Q,{\cal D}]=\nn
&&-\frac{g^2}{2}\int_\q\sum_{k,n} \left(\Delta^>(E'_{kn},\q)  [{\cal S}_{\q\cdot\hat{\boldsymbol r}},{\cal P}_n{\cal S}_{\q\cdot\hat{\boldsymbol r}}{\cal P}_k{\cal D}(t)]+\Delta^>(E'_{nk},\q)[{\cal D}(t){\cal P}_n{\cal S}_{\q\cdot\hat{\boldsymbol r}}{\cal P}_k,{\cal S}_{\q\cdot\hat{\boldsymbol r}}]\right)\nonumber\\
&&+i g^2\int_\q\sum_{k,n}{\rm P}\int_{q_0}\Delta^>(q_0,\q)\left( \frac{1}{q_0-E'_{kn}}-\frac{1}{q_0}\right) [{\cal S}_{\q\cdot\hat{\boldsymbol r}},{\cal P}_n{\cal S}_{\q\cdot\hat{\boldsymbol r}}{\cal P}_k{\cal D}(t)]\nn
&&-i g^2\int_\q\sum_{k,n}{\rm P}\int_{q_0}\Delta^>(q_0,\q)\left( \frac{1}{q_0+E'_{kn}}-\frac{1}{q_0}\right) [{\cal D}(t){\cal P}_n{\cal S}_{\q\cdot\hat{\boldsymbol r}}{\cal P}_k,{\cal S}_{\q\cdot\hat{\boldsymbol r}}],
\label{eq:evolt3b}
\end{eqnarray}
where  the energies $E'_n$ are the eigenvalues of $H'_Q$. A this point, it is convenient to consider the explicit matrix elements of ${\cal D}$ and write the equation  in a Liouvillian form
\beq\label{Liouville1}
\frac{\rmd{\cal D}_{ij}}{dt}+i E_{ij} {\cal D}_{ij}={\cal L}_{ij,kl} {\cal D}_{kl}.
\eeq
In order to simplify the writing, we set $\bra{i}{\cal S}_{\q\cdot\hat{\boldsymbol r}}\ket{j}\to {\cal S}_{ij}$ in the following. We then obtain
\beq\label{Liouvillianijkl}
&&{\cal L}_{ij,kl}=-\frac{g^2}{2} \left({\cal S}_{in}{\cal S}_{nk}\delta_{jl}\,\Delta^>(E'_{kn},\q) - {\cal S}_{ik}{\cal S}_{lj}\,\Delta^>(E'_{ki},\q)\right) \nn
&&\qquad \quad -\frac{g^2}{2} \left({\cal S}_{ln}{\cal S}_{nj}\delta_{ik}\, \Delta^>(E'_{ln},\q) -{\cal S}_{ik}{\cal S}_{lj} \,\Delta^>(E'_{lj},\q)\right)\nn
&&\qquad\quad +ig^2\left({\cal S}_{in}{\cal S}_{nk}\delta_{jl}\,{\rm P}'\int_{q_0}\frac{\Delta^>(q_0,\q)}{q_0-E'_{kn}} - {\cal S}_{ik}{\cal S}_{lj}\,{\rm P}'\int_{q_0}\frac{\Delta^>(q_0,\q)}{q_0-E'_{ki}}\right) \nn
&&\qquad\quad -ig^2\left({\cal S}_{ln}{\cal S}_{nj}\delta_{ik}\, {\rm P}'\int_{q_0}\frac{\Delta^>(q_0,\q)}{q_0+E'_{nl}} -{\cal S}_{ik}{\cal S}_{lj} \,{\rm P}'\int_{q_0}\frac{\Delta^>(q_0,\q)}{q_0+E'_{jl}}\right).\nn
\eeq
In this equation, ${\rm P}'$ denotes the principal part of the integral from which the contribution $1/q_0$ is subtracted (cf. Eq.~(\ref{eq:evolt3b})).

Assuming that the Liouvillian in the right hand side of Eq.~(\ref{Liouville1}) can be treated as a perturbation, we expect the effect of this perturbation to be dominant when it connects pairs of states with comparable energy differences, that is, when $|E_{ij}|\simeq |E_{kl}|$. In particular, one expects the   diagonal elements of the density matrix, i.e. the occupation probabilities of the various levels, for which $E_{ij}=0$ to decouple from the non diagonal ones. \footnote{A more formal discussion of this issue is presented in Appendix~\ref{secular}.} 
We now restrict ourselves to these diagonal matrix elements, and ignore possible degeneracies. It is easy to verify that the principal values then cancel. We  get, for the case $i\ne k$,
\beq
{\cal L}_{ii,kk}=g^2|{\cal S}_{ik}|^2 \Delta^>(E_{ki},\q).
\eeq
This is nothing but the decay rate $\Gamma_{k\to i}$ calculated according to Fermi's golden rule. That is,
\beq\label{rate1}
\Gamma_{k\to i}=g^2\int_\q|\bra{i}{\cal S}_{\q\cdot\hat{\boldsymbol r}}\ket{k}|^2 \Delta^>(E_{ki},\q), 
\eeq
where $\Delta^>(E_{ki},\q)$ plays the role of the density of available states for the transition $k\to i$. 
Similarly, for the case $i=k$ we get
\beq
{\cal L}_{ii,ii}=\sum_{j\ne i}\Gamma_{i\to j}.
\eeq

At this point we denote by $p_n(t)=\bra{n}{\cal D}(t)\ket{n}$ the probability to find the system in the eigenstate $n$ of $H'_Q$.  The equation (\ref{Liouville1}) yields then
\beq\label{dpndt}
\frac{\rmd p_n}{\rmd t}= \sum_k\left( p_k \Gamma_{k\to n}-p_n \Gamma_{n\to k}\right).
\eeq
From the property
\begin{equation}
\Delta^>(E_{kn},{\bf q})=e^{-\frac{E_{nk}}{T}}\Delta^>(E_{nk},{\bf q}),
\label{eq:flucdiss}
\end{equation}
and Eq.~(\ref{rate1}), it follows that 
\beq
\Gamma_{k\to n}=\rme^{-E_{nk}/T}\,\Gamma_{n\to k}.
\eeq
Thus, in equilibrium where $\rmd p_n/\rmd t=0$, the detailed balance relation $p_k\Gamma_{k\to n}=p_n\Gamma_{n\to k}$ implies 
\beq
\frac{p_k}{p_n}=\rme^{-(E_k-E_n)/T},
\eeq
that is $p_n\propto \rme^{-E_n/T}$. In other words, the system reaches thermal equilibrium at the temperature of the plasma.\\

In order to see globally how the equilibrium is achieved, we look at the free energy $F$, defined in terms of the density matrix as in equilibrium, viz.
\beq
F={\rm Tr} H'_Q{\cal D}+T{\rm Tr}{\cal D}\ln{\cal D}=\sum_n(E_n+T\log p_n)p_n,
\eeq
where in the last equality, the states $n$ are the eigenstates of $H'_Q$. 
Taking the time derivative of this equation, and using Eqs.~(\ref{dpndt}) and (\ref{eq:flucdiss}), we get
\beq
\frac{\rmd F}{\rmd t}&=&-\sum_{nk}(E_n+T\log p_n)(p_n-e^{-\frac{E_{nk}}{T}}p_k)\Gamma_{n\to k}\nn
&=&-\frac{T}{2}\sum_{nk}(p_n-e^{-\frac{E_{nk}}{T}}p_k)\log\left(\frac{p_n}{p_ke^{-\frac{E_{nk}}{T}}}\right)\Gamma_{n\to k}.
\eeq
Since all the terms in the sum are positive, this equation shows that  the free energy  is a monotonously decreasing function of time (at least in the large time limit).  An alternative and more formal proof can be obtained by using  \textit{Lemma} 1 of Ref.~\cite{:/content/aip/journal/jmp/19/5/10.1063/1.523789}. Note that if we use  this evolution equation to compute the derivatives of $S$ and $E$ separately, one obtains expressions that do not have necessarily a well defined sign. 

We may estimate the rate of  change in the free energy, using a similar argument as that used for the entropy.  We get
\beq\label{rateF}
\frac{1}{F}\frac{\rmd F}{\rmd t}&\approx& -4g^2\int_\q\Delta^>(\Delta E,{\bf q})\left(\sin\left(\frac{{\bf q}{\bf a_0}}{2}\right)\right)^2\nn
&\approx&2 \Gamma(\Delta E, \a_0),
\eeq
where $\Delta E$ is a quantity that represents an average value for the binding energy differences. In the last line $\Gamma(\Delta E, \a_0)$ is a damping rate that summarizes the effect of the collisions. This can be viewed as a definition that generalizes that of $\Gamma(\a_0)$ to include an energy dependence (i.e. $\Gamma(\a_0)=\Gamma(\Delta E=0, \a_0)$, see last section). Eq.~(\ref{rateF})  is essentially the same as that giving the entropy increase, Eq.~(\ref{eq:Sch}), with $\Delta^>(0,{\bf q})$ replaced by $\Delta^>(\Delta E,{\bf q})$.  We shall discuss in the last section how large is the correction due to this energy dependence. \\

\section{Entropy and free energy in a non-Abelian theory}
\label{sec:EFQCD}
In this section we repeat the analysis of the previous section in the case of QCD. The generalization is straightforward except for obvious complications related to the color algebra, and the existence of several components of the density matrix.

\subsection{Entropy increase}

As in the abelian case, in the high temperature limit, the evolution equations for the reduced density matrix  are a set of Lindblad equations, in which the effect of collisions is taken into account via an imaginary potential.  We shall verify that the entropy increases monotonously as the equilibrium is approached. 

Following the same reasoning as in Sect.~\ref{sec:QEDhighT} we get the following simplified expressions for the various operators ${\cal L}$: 
\begin{equation}\label{eq84}
{\cal L}^{\rm ss}D_{\rm s}=-\frac{C_F}{2}\int_\q[L_{\q}L_{\q},D_{\rm s}]-\frac{C_F}{2}\int_\q\{L_{\q}L_{\q},D_{\rm s}\}\,,
\end{equation}
\begin{equation}\label{eq85}
{\cal L}^{\rm so} D_{\rm o}=C_F\int_\q L_{\q}D_{\rm o}L_{\q}\,,
\end{equation}
\begin{equation}\label{eq86}
{\cal L}^ {\rm os}D_{\rm s}=\frac{1}{2N_c}\int_\q L_{\q}D_{\rm s}L_{\bf q}\,,
\end{equation}
and 
\begin{equation}\label{eq87}
{\cal L}^{\rm oo}_1D_{\rm o}=-\frac{1}{4N_c}\int_\q[L_{\q}L_{\q},D_{\rm o}]-\frac{1}{4N_c}\int_\q\{L_{\q}L_{\q},D_{\rm o}\}\,.
\end{equation}
\begin{equation}
{\cal L}^{\rm oo}_2 D_{\rm o}=\frac{(N_c^2-4)}{4N_c}\int_\q(L_{\q}D_{\rm o} L_{\q}-\frac{1}{2}\{L_{\q}L_{\q},D_{\rm o}\})\,.
\end{equation}
\begin{equation}
{\cal L}^{\rm oo}_3D_{\rm o}=\frac{N_c}{4}\int_\q(\bar L_{\q}D_{\rm o} \bar L_{\q}-\frac{1}{2}\{\bar L_{\q} \bar L_{\q},D_{\rm o}\})\,.
\end{equation}
In the previous equations we have used $L_{\q}$ as defined in Eq. (\ref{eq:defCq}) while
\begin{equation}
\bar L_{\q\cdot{\boldsymbol r}}\equiv 2g\sqrt{\Delta^>(0,{\q})}\cos\left({\q}\cdot{{\boldsymbol r}}/{2}\right)\,.
\end{equation}

In the QCD case the entropy can be written
\begin{equation}
S=-{\rm Tr}(D_{\rm s}\log D_{\rm s})-(N_c^2-1){\rm Tr}(D_{\rm o}\log D_{\rm o})\,,
\end{equation}
where the factor $N_c^2-1$ is due to the normalization chosen in Eq.~(\ref{eq:rhodecom1}). 
By using the explicit expression of the operators ${\cal L}^{ij}$ just given above, we  can write the derivative of the entropy as
\begin{eqnarray}
&&\frac{\rmd S}{\rmd t}=-{\rm Tr}(({\cal L}^{\rm ss}D_{\rm s}+{\cal L}^{\rm so}D_{\rm o})\log D_{\rm s})-(N_c^2-1){\rm Tr}(({\cal L}^{\rm os}D_{\rm s}+{\cal L}^{\rm oo}_1D_{\rm o})\log D_{\rm o})\nonumber \\
&&-(N_c^2-1){\rm Tr}({\cal L}^{\rm oo}_2 D_{\rm o}\log D_{\rm o})-(N_c^2-1){\rm Tr}({\cal L}^{\rm oo}_3 D_{\rm o}\log D_{\rm o})\,.
\label{eq:dSdt}
\end{eqnarray}
Using exactly the same reasoning as in Sect.~\ref{sec:QEDhighT} we can then show that the second line of Eq.~(\ref{eq:dSdt}) is positive, i.e., 
\begin{equation}
-(N_c^2-1){\rm Tr}\left[ ({\cal L}^{\rm oo}_2 D_{\rm o}+ {\cal L}^{\rm oo}_3 D_{\rm o} )\log D_{\rm o}\right]\ge 0.
\end{equation}
The first line of Eq.~(\ref{eq:dSdt}) introduces an additional complication, in praticular because it mixes $D_{\rm s}$ and $D_{\rm o}$. At this point, we use a (time-dependent) basis in which  $D_{\rm s}$ and $D_{\rm o}$ are diagonal, that is we set
\begin{equation}\label{DsDospectral}
D_{\rm s}=\sum_n p^{\rm s}_n|{\rm s},n\rangle\langle {\rm s},n|,\qquad 
D_{\rm o}=\sum_m p^{\rm o}_m|{\rm o},m\rangle\langle {\rm o},m|, 
\end{equation}
where, in the last expression,  $|{\rm o},m\rangle\langle {\rm o},m|$ actually stands for 
\beq
|{\rm o},m\rangle\langle {\rm o},m|=\frac{1}{N_c^2-1}\sum_C |{\rm o}^C,m\rangle\langle {\rm o}^C,m|.
\eeq
Then, a simple calculation allows us to write  the first line of the right hand side of Eq.~(\ref{eq:dSdt}) as
\begin{eqnarray}
&&4g^2C_F\sum_{nm}(p_n^{\rm s}-p_m^{\rm o})\log\left(\frac{p_n^s}{p_m^{\rm o}}\right)\int_\q|\langle {\rm s},n|L_{\bf q}|{\rm o},m\rangle|^2\geq 0\,.
\end{eqnarray}
In conclusion, all the terms contributing to the derivative of the entropy are positive. This implies that in the regime where the temperature of the quark-gluon plasma is large in comparison to the typical binding energies, the equations for the reduced density matrix yield a monotonous increases of the entropy as the quarkonium approaches thermal equlibrium. The rate  of entropy change can be estimated in the same way as we did for the abelian case  in Sect.~\ref{sec:QEDhighT}.

\subsection{Free energy minimization}

With consider now the regime of moderate temperatures, and will proceed to the calculation of the free energy. We shall first write the necessary rate equations describing the evolution of the populations of the various states. Although many of these states belong to a continuum, we write the summations over states as discrete sums, as in Eq.~(\ref{DsDospectral}), instead of integrations, since we focus here on the general structure of the equations. The contimuum states will be explicitly dealt with in the examples treated in the next section. 

The probabilities $p_n^{\rm s}$ and $p_m^{\rm o}$ fulfil the following evolution equations
\begin{equation}
\frac{\rmd p_n^s}{\rmd t}=g^2C_F\sum_m\left(p_m^{\rm o}-p_n^{\rm s} e^{-\frac{E_m^o-E_n^s}{T}}\right)\int_\q\Delta^>(E_m^o-E_n^s,{\q})|\langle {\rm s},n|{\cal S}_{\q\cdot\hat{\boldsymbol r}} |{\rm o},m\rangle|^2\,,
\label{eq:evops}
\end{equation}
and
\begin{eqnarray}
&&\frac{\rmd p_m^o}{\rmd t}=-\frac{g^2}{2N_c}\sum_n\left(p_m^o-p_n^se^{-\frac{E_m^{\rm o}-E_n^{\rm s}}{T}}\right)\int_\q \Delta^>(E_m^{\rm o}-E_n^{\rm s},{\q})|\langle {\rm s},n|{\cal S}_{\q\cdot\hat{\boldsymbol r}} |{\rm o},m\rangle|^2\nonumber\\
&&-\frac{g^2(N_c^2-4)}{4N_c}\sum_k\left(p_m^{\rm o}-p_k^{\rm o}\rme^{-\frac{E_m^{\rm o}-E_k^{\rm o}}{T}}\right)\int_\q \Delta^>(E_m^{\rm o}-E_k^{\rm o},{\q})|\langle {\rm o},m|{\cal S}_{\q\cdot\hat{\boldsymbol r}} |{\rm o},k\rangle|^2\nonumber\\
&&-\frac{g^2N_c}{4}\sum_k\left(p_m^{\rm o}-p_k^{\rm o}\rme^{-\frac{E_m^{\rm o}-E_k^{\rm o}}{T}}\right)\int_\q\Delta^>(E_m^{\rm o}-E_k^{\rm o},{\q})|\langle {\rm o},m| {\cal C}_{\q\cdot\hat{\boldsymbol r}} |{\rm o},k\rangle|^2\,.\nn
\label{eq:evopo}
\end{eqnarray}
Note that in order to obtain Eq.~(\ref{eq:evops}),  we had to combine Eq.~(\ref{eq84}) and (\ref{eq85}), while the first of Eqs.~(\ref{eq:evopo}) involves both Eq.~(\ref{eq86}) and (\ref{eq87}). The structure here is very much like what occurs in the entropy calculation, Eq.~(\ref{eq:dSdt}).

We may define, in agreement with Fermi's golden rule (see Eq.~(\ref{rate1})),
\beq
\Gamma_{{\rm o},m\to{\rm s},n}=\frac{g^2}{2N_c}\int_\q \Delta^>(E_m^{\rm o}-E_n^{\rm s},{\q})|\langle {\rm s},n|{\cal S}_{\q\cdot\hat{\boldsymbol r}} |{\rm o},m\rangle|^2,\nn
\Gamma_{{\rm s},n\to{\rm o},m}=g^2 C_F\int_\q \Delta^>(E_n^{\rm s}-E_m^{\rm o},{\q})|\langle {\rm s},n|{\cal S}_{\q\cdot\hat{\boldsymbol r}} |{\rm o},m\rangle|^2.
\eeq
The first equation gives the transition rate $\Gamma_{{\rm o}\to{\rm s}}$ from one particular member of an octet state to a singlet state (the factor $1/(2N_c)$ follows from Eq.~(\ref{colmelnA})). In the second equation, giving $\Gamma_{{\rm s}\to{\rm o}}$, all members of the considered octet are summed over (producing a factor $N_c^2-1)$). Similarly, we have, for the octet to octet transitions
\beq
&&\Gamma_{{\rm o},m\to{\rm o},k}^{(2)}=\frac{g^2(N_c^2-4)}{4N_c}\int_\q \Delta^>(E_m^{\rm o}-E_k^{\rm o},{\q})|\langle {\rm o},m|{\cal S}_{\q\cdot\hat{\boldsymbol r}} |{\rm o},k\rangle|^2,\nn
&&\Gamma_{{\rm o},m\to{\rm o},k}^{(2)}=\frac{g^2 N_c}{4}\int_\q \Delta^>(E_m^{\rm o}-E_k^{\rm o},{\q})|\langle {\rm o},m|{\cal C}_{\q\cdot\hat{\boldsymbol r}} |{\rm o},k\rangle|^2.
\eeq
All these transition rates are those which control the evolution of the populations, according to the equations written above.


With these ingredients we can compute the evolution of the free energy. Expanding on the basis of the eigenstates of $H_Q'$, and dropping the $'$ on the energies in order to simplify the notation, one gets
\begin{equation}\label{freeenergyQCD}
F=\sum_n(E_n^s p_n^s+Tp_n^s\log p_n^s)+(N_c^2-1)\sum_m(E_m^o p_m^o+Tp_m^o\log p_m^o)\,.
\end{equation}
Using analogous manipulations as the ones we used in section \ref{sec:EQED} we get
\begin{eqnarray}
\frac{\rmd F}{\rmd t}=&-&T(N_c^2-1)\sum_{nm}\log\left(\frac{p_m^o}{p_n^se^{-\frac{E_m^o-E_n^s}{T}}}\right)\left(p_m^o-p_n^se^{-\frac{E_m^o-E_n^s}{T}}\right)\Gamma_{{\rm o},m\to{\rm s},n}\nonumber \\
&-&\frac{ T}{2}\sum_{mk}\log\left(\frac{p_m^o}{p_k^oe^{-\frac{E_m^o-E_k^o}{T}}}\right)\left(p_m^o-p_k^oe^{-\frac{E_m^o-E_k^o}{T}}\right)\Gamma_{{\rm o},m\to{\rm o},k}^{(2)}\nonumber\\
&-&\frac{T}{2}\sum_{mk}\log\left(\frac{p_m^o}{p_k^oe^{-\frac{E_m^o-E_k^o}{T}}}\right)\left(p_m^o-p_k^oe^{-\frac{E_m^o-E_k^o}{T}}\right)\Gamma_{{\rm o},m\to{\rm o},k}^{(3)}.\nn
\end{eqnarray}
Again the physical interpretation is straightforward: each of the elementary transitions makes the free energy decrease separately, in a way that is very similar to what happens in the Abelian limit.


\section{Some illustrative calculations}
\label{sec:illustrations}

In this last section, we present results of some numerical solutions of the equations for the density matrix in simplified situations. We emphasize that our main goal here is to illustrate some of the concepts that we have introduced. Thus, although the numbers are adjusted to bottomonium physics, we make no attempt to a complete phenemonological description.  The first example that we treat is that of an infinitely massive quark-antiquark pair. This provides a simple illustration of  the role of the energy dependence of the imaginary potential, as well as a quantitative indication of the magnitude of the effect. In principle, such a setting corresponds to that used in lattice QCD calculations, and we briefly compare with relevant lattice results. The next example involves a simplified picture of a bottomonium in a plasma, with a single bound state in the singlet channel, and octets states involving the free quark and antiquark. In this case, rate equations are complemented by a Langevin equation describing the Brownian motion of the heavy quark and antiquark in the plasma. 

\subsection{Infinitely massive quark-antiquark pair}

The physics of singlet to octet transitions is best analyzed by ignoring completely the motion of the heavy particles and focussing on their color degrees of freedom alone. This is what we do in this subsection. 

We consider an infinitely massive quark-antiquark pair, and assume that it can exist in two color states, a singlet (s) and $N_c^2-1$ degenerate octet (o) states. There are no continuum states, so that the problem reduces to that a two level system. The partition function reads
\beq
Z=\rme^{-\frac{V_{\rm s}}{T}}+(N_c^2-1)\rme^{-\frac{V_{\rm o}}{T}},
\eeq
where $V_{\rm s}({\boldsymbol r})$ and $V_{\rm o}({\boldsymbol r})$ denote the energies of the pair in a singlet and an octet state, respectively. Since the particles do not move, they have no kinetic energy, and the energy of the pair is just the potential energy, which depends on the distance ${\boldsymbol r}$ between the quark and the antiquark. We assume that $V_{\rm s}<V_{\rm o}$, and set $\Delta V=V_{\rm o}-V_{\rm s}$. The free energy is given by $F=E-TS=-T\ln Z$, with $E$ the internal energy and $S$ the entropy. The latter  can be deduced from $F$ by using the thermodynamic relation  $S=-\del F/\del T$. In the low temperature limit,
\beq
F\approx V_{\rm s}-T(N_c^2-1) \rme^{-\frac{\Delta V}{T}}.
\eeq
In this regime, the dominant contribution to the free energy is the energy of the ground state $V_{\rm s}$, the correction from the octet excited states being  exponentially small, $\propto T\rme^{-\frac{\Delta V}{T}}$. The internal energy and entropy are given by 
\beq
E=V_{\rm s}+(N_c^2-1)\Delta V \, \rme^{-\frac{\Delta V}{T}},\qquad S=(N_c^2-1) \rme^{-\frac{\Delta V}{T}}\left[ 1+\frac{\Delta V }{T} \right].
\eeq
In the opposite limit of high temperature, the free energy is entirely dominated by the entropy. A simple calculation yields indeed
\beq
F=-T\ln N_c^2+\frac{V_{\rm s}+(N_c^2-1) V_{\rm o}}{N_c^2},\qquad S=\ln N_c^2.
\eeq
The factor $N_c^2$ is just the total number of available states, one singlet and $(N_c^2-1)$ octet. All are present in equilibrium with the same probability. The Boltzmann factors in the partition function can all be approximated by unity, and the internal energy is simply given by
\beq
E=\frac{1}{N_c^2} \,V_{\rm s}+\left(1-\frac{1}{N_c^2}  \right)V_{\rm o}.
\eeq
It is independent of the temperature. 

Let  $p_{\rm s}$ and $p_{\rm o}$ be the probabilities to  find the system respectively in the singlet or a given octet state. In the infinite mass limit,  these are simply the diagonal elements of the density matrix (cf. Eq.~(\ref{DlargeM})), i.e.,   $p_{\rm s}=D_{\rm s}({\boldsymbol r})$ and $p_{\rm o}=D_{\rm o}({\boldsymbol r})$. In equilibrium, we have
\beq
 p_{\rm s}=\frac{\rme^{-V_{\rm s}/T}}{Z},\qquad p_{\rm o}=\frac{\rme^{-V_{\rm o}/T}}{Z}, \qquad p_{\rm s}+(N_c^2-1) p_{\rm o}=1.
\label{eq:staticeq}
 \eeq
 These probabilities depend only on the ratio $\Delta V/T$, which controls the transition between the low and the high temperature regimes:
 \beq\label{probaequil}
 \frac{p_{\rm s}}{p_{\rm o}}=\rme^{\Delta V/T}, \qquad p_{\rm s}=\frac{1}{1+(N_c^2-1) \rme^{-\Delta V/T}}.
 \eeq
 At low temperature, $T\ll \Delta V$, and $p_{\rm s}\lesssim 1$. At high temperature, $T\gg \Delta V$,  $p_{\rm s}\simeq p_{\rm o}\simeq 1/N_c^2$. In the numerical calculations, we use 
\begin{equation}
\Delta V(r)=\frac{N_c\alpha_s(1/r)e^{-m_D r}}{2r}\,,
\end{equation}
where $m_D$ is the HTL Debye mass calculated with a running coupling at the scale $2\pi T$.
\\

We are interested in the dynamics of the approach to the equilibrium state. As we have seen in the previous sections, this is controlled by a rate equation, generically of the form 
\beq\label{rate0}
\frac{\rmd p_{\rm s}}{\rmd t}=(N_c^2-1) p_{\rm o}\Gamma_{{\rm o}\to {\rm s}}-p_{\rm s}\Gamma_{{\rm s}\to {\rm o}}, 
\eeq
where $\Gamma_{{\rm o}\to {\rm s}}$ denotes the transition rate from any one of $N_c^2-1$ degenerate octet states, and similarly for $\Gamma_{{\rm s}\to {\rm o}}$.
In a stationary state, the rate equation $\rmd p_{\rm s}/\rmd t=0$ yields the detailed balance condition
\beq\label{detbal}
\frac{p_{\rm s}}{p_{\rm o}}=\frac{(N_c^2-1) \Gamma_{{\rm o}\to {\rm s}}}{\Gamma_{{\rm s}\to {\rm o}}}.
\eeq
This is to be compared to the result that we expect when the stationary state is the state of  thermal equilibrium (cf. Eq.~(\ref{probaequil}))
\begin{equation}
\frac{p_{\rm s}({\bf r})}{p_{\rm o}({\bf r})}= \rme^{\frac{\Delta V({\bf r})}{T}} .
\label{eq:ratioTside}
\end{equation}
By comparing the two equations (\ref{eq:ratioTside}) and (\ref{detbal}) one gets the relation
\beq\label{FD2}
\Gamma_{{\rm s}\to {\rm o}}\,\rme^{-V_{\rm s}/T}=(N_c^2-1)\Gamma_{{\rm o}\to {\rm s}}\,\rme^{-V_{\rm o}/T}.
\eeq
This relation is satisfied by the rates that we have obtained in the previous section. It implies in particular that their energy dependence needs to be taken into account when the temperature is of the order of magnitude of $\Delta V$: in that case the static imaginary potential is not sufficient to fully account for the effects of collisions. We return to this issue shortly.

\begin{figure}
\begin{center}
\includegraphics[scale=0.5]{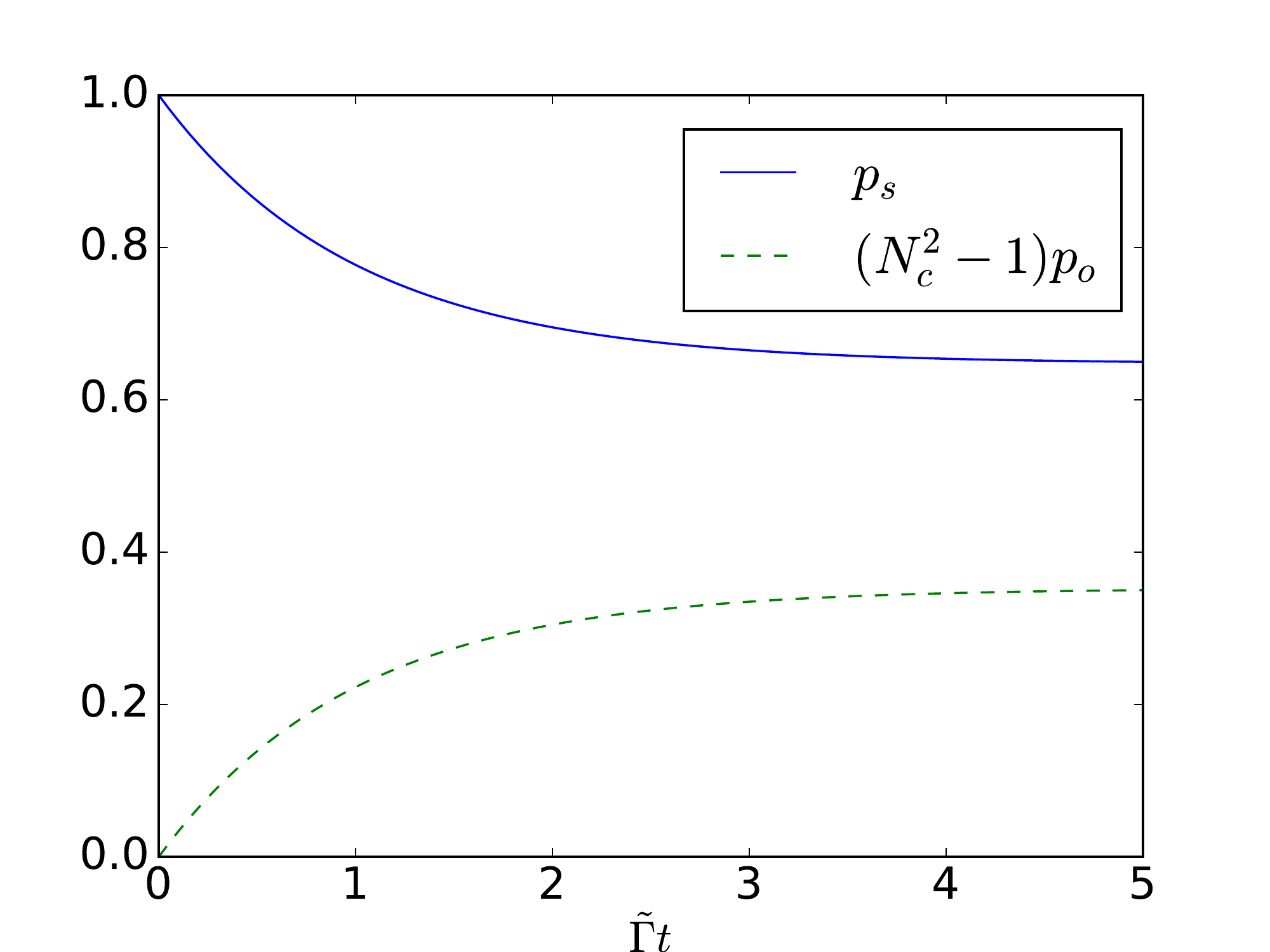}\\
\includegraphics[scale=0.4]{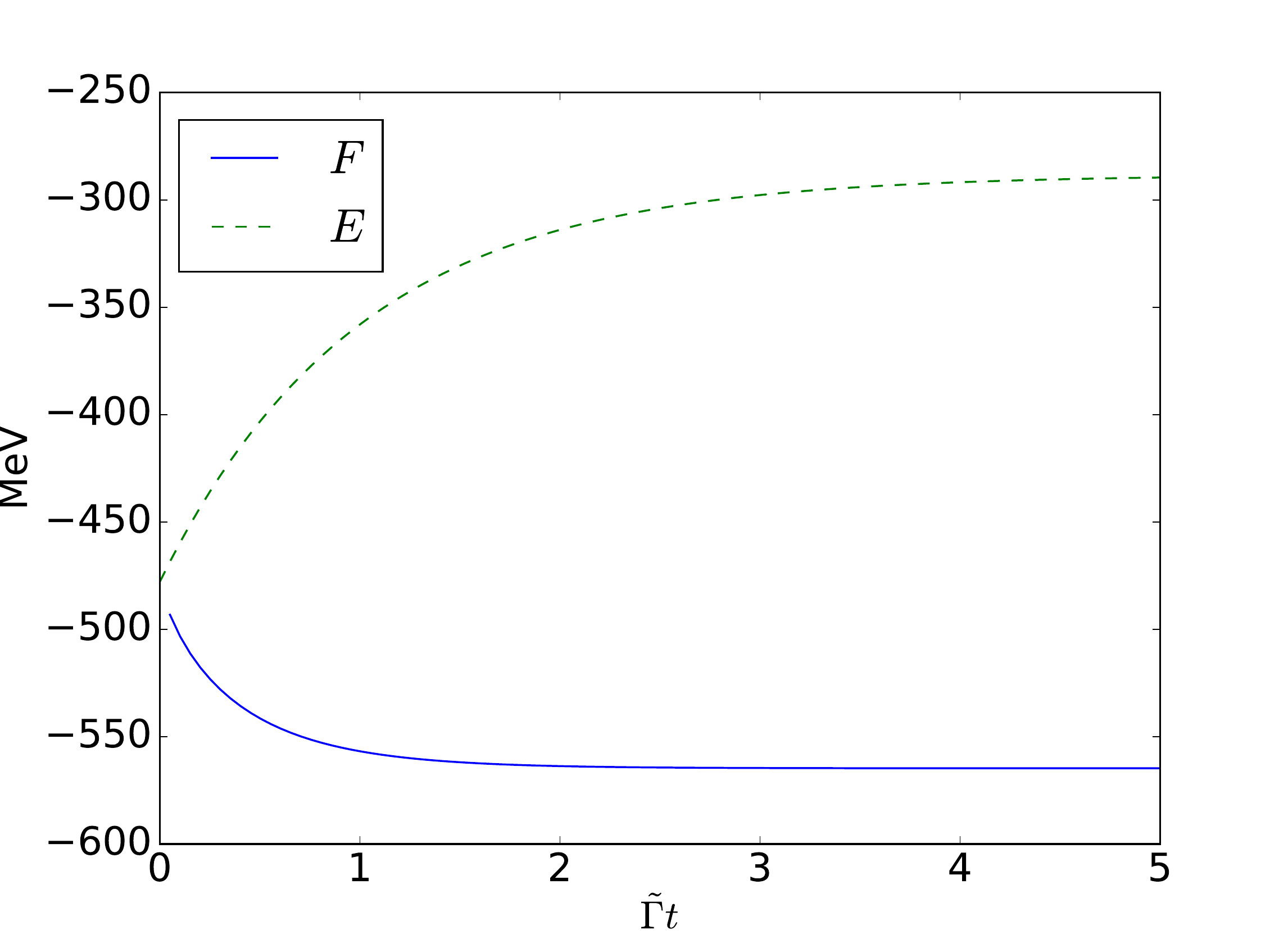}
\caption{Top: Time evolution of the probabilities $p_{\rm s}$ and $p_{\rm o}$, assuming $p_{\rm s}(0)=1$ and $r=0.15\,\rm{fm}$,   corresponding approximately to the Bohr radius $a_0$ of the singlet bound state. Bottom: Free energy computed in the same conditions.  The temperature of the plasma is $T=200 MeV$. }\label{fig:pstatic}
\end{center}
\end{figure}

The evolution equation (\ref{rate0}) has the following solution,  for arbitrary initial conditions,
\begin{equation}
p_{\rm s}(t)=p_{\rm s}^{\rm eq}(1-e^{-\tilde{\Gamma}t})+p_{\rm s}(0)e^{-\tilde{\Gamma}t}\,,
\end{equation}
and
\begin{equation}
p_{\rm o}(t)=p_{\rm o}^{\rm eq}\left[1+\frac{\rme^{\Delta V/T}\rme^{-\tilde{\Gamma}t}}{(N_c^2-1)}\right]-p_{\rm s}(0)\,\frac{\rme^{-\tilde{\Gamma}t}}{N_c^2-1},
\end{equation}
where $p_{\rm s}^{\rm eq}$ and $p_{\rm o}^{\rm eq}$ are the equilibrium values, given in Eq.~(\ref{eq:staticeq}), and $\tilde{\Gamma}$ is an effective rate defined as
\begin{equation}
\tilde{\Gamma}=\Gamma_{{\rm s}\to {\rm o}}\left(1+\frac{\rme^{\Delta V/T}}{N_c^2-1}\right)\,.
\end{equation}
The solution $p_{\rm s} (t)$ and $p_{\rm o}(t)$ obtained for  $p_{\rm s}(0)=1$ and $r=0.15\,\rm{fm}$ are plotted  in Fig.~\ref{fig:pstatic}, together with the free energy calculated from Eq.~(\ref{freeenergyQCD}). For both the survival probability $p_{\rm s}(t)$  and the free energy,   the effective rate $\tilde{\Gamma}({\boldsymbol r})$ determines the time scale that controls the approach to equilibrium. 

In order to quantify the importance of the energy dependence of the rates, we use the  explicit results that we have obtained in the previous section.  
From Eq.~(\ref{eq:evops}),  we get
\begin{equation}\label{rate2}
\frac{dp_{\rm s}}{dt}=g^2C_F\left(p_{\rm o}-e^{-\frac{\Delta V}{T}}p_{\rm s}\right)\int\frac{\,d^3q}{(2\pi)^3}\Delta^>(\Delta V,{\bf q})\left({\cal S}_{\q\cdot\hat{\boldsymbol r}} \right)^2.
\end{equation}
This equation is identical to  Eq.~(\ref{rate0}), with now the following explicit expressions for the rates
\beq\label{ratesGammasos}
&&\Gamma_{{\rm o}\to{\rm s}}=\frac{g^2}{2N_c}\int_\q \Delta^>(\Delta V,\q)|{\cal S}_{\q\cdot\hat{\boldsymbol r}} |^2,\nn
&&\Gamma_{{\rm s}\to{\rm o}}={g^2}C_F\int_\q \Delta^<(\Delta V,\q)|{\cal S}_{\q\cdot\hat{\boldsymbol r}} |^2\nn
&&\qquad\;\; = {g^2}C_F\rme^{-\frac{\Delta V}{T}}\int_\q \Delta^>(\Delta V,\q)|{\cal S}_{\q\cdot\hat{\boldsymbol r}} |^2.
\eeq
It is easily verified that these rates satisfy Eq.~(\ref{FD2}), as stated above.

In the static limit, and at high temperature, one can express the survival probability of the singlet state in terms of an imaginary  potential. We have
\beq\label{Gammarms}
\Gamma_{\rm s}({\boldsymbol r})&\equiv&C_F \left( W({\boldsymbol r})-W(0)\right)=C_F \int_q W(\q)\left( \rme^{i\q\cdot{\boldsymbol r}}-1  \right)\nn
&=&\frac{g^2C_F}{2}\int_q \Delta^<(0,{\bf q})\left({\cal S}_{\q\cdot\hat{\boldsymbol r}} \right)^2\nn
&=& \frac{1}{2} \Gamma_{{\rm s}\to{\rm o}},
\eeq
where we have used $W(\q)=- g^2\Delta^<(0,{\bf q})$, and in the last line $\Gamma_{{\rm s}\to{\rm o}}$ is given by Eq.~(\ref{ratesGammasos}) in which one sets $\Delta V=0$. This relation suggests the following  definition of a generalized, energy dependent, ``potential'' $W(\omega,{\boldsymbol r})$, viz. 
\beq
W(\omega,{\boldsymbol r})-W(\omega,0)&\equiv&\frac{g^2}{2} \int\frac{\,d^3q}{(2\pi)^3}\Delta^<(\omega,{\bf q})\left({\cal S}_{\q\cdot\hat{\boldsymbol r}} \right)^2,\nn
&=&\frac{g^2}{2}\int\frac{\,d^3q}{(2\pi)^3}N(\omega)\sigma(\omega,{\bf q})\left({\cal S}_{\q\cdot\hat{\boldsymbol r}} \right)^2\nn
&=&g^2N(\omega)\int\frac{\,d^3q}{(2\pi)^3}\sigma(\omega,{\bf q})\left(1-\rme^{i\q\cdot\hat{\boldsymbol r}}   \right)
\label{eq:ImVresum}
\eeq
where $\sigma(\omega,\q) =\Delta^>(\omega,{\bf q})-\Delta^<(\omega,{\bf q})$ is the (longitudinal) gluon spectral function, and we have used the relation 
\beq\label{relationsdeltasigma}
\Delta^<(\omega,{\bf q})=\sigma(\omega,\q) N(\omega),\qquad N(\omega)=\frac{1}{\rme^{\omega/T}-1}.
\eeq
Note that as $\omega\to 0$,  $W(\omega,{\boldsymbol r})$ reduces to $W({\boldsymbol r})$ since, in this limit, $N(\omega)\sim T/\omega$, and $T\sigma(\omega,\q)/\omega\to \Delta^<(0,\q)$. 

It is perhaps useful to recall here a few basic properties of the gluon spectral function $\sigma(\omega,\q)$.  To be specific, we shall rely on the HTL approximation, for which an explicit expression is known (see e.g. \cite{Beraudo:2007ky})\footnote{Note that all the numerical calculations presented in this paper use this approximation. Note also that we shall be using this approximation beyond its strict regime of validity, which requires $\omega, q\ll T$. }.  At fixed momentum, $\sigma(\omega,\q)$ is an increasing function of the energy (linear at small energy), in the space-like domain $|\omega|<|\q|$. For $|\omega|>|\q|$ it vanishes, except for an isolated delta-function contribution corresponding to the plasmon excitation at $\omega_\q$ ($\omega_\q^2\simeq \omega_{\rm pl}^2+6q^2/5$, with $\omega_{\rm pl}=m_D/\sqrt{3}$), which exists only for $|\q|\lesssim m_D$.  The specific contribution of the plasmon to Eq.~(\ref{eq:ImVresum}) will be ignored in this paper.\footnote{It should of course be included in a more quantitative study. It represents a process analogous to that of gluon dissociation involving the transverse modes of the gluon \cite{Bhanot:1979vb,Brambilla:2011sg,Brezinski:2011ju}. The collective plasmon exists only at small momentum $q\lesssim m_D$, and its contribution to $W(\omega,{\boldsymbol r})-W(\omega,0)$ is expected to be modest in the region of interest, and taking it into account would not alter the main conclusions of this section. }
For $\omega=0$,  we know $\sigma(\omega,\q)$, and hence $\Delta^<(0,\q)$, analytically. This is
\beq\label{eq.122}
\Delta^<(0,\q)=\frac{\pi m_D^2 T}{|\q|(\q^2+m_D^2)^2},
\eeq
so that \cite {Laine:2006ns}
\beq\label{GammaHTL}
\Gamma({\boldsymbol r})=W({\boldsymbol r})-W(0)=\frac{g^2T}{2\pi } \int_0^\infty \rmd x  \frac{x}{(x^2+1)^2} \left[1-\frac{\sin(x rm_D)}{x rm_D}   \right].
\eeq

When the energy is non vanishing, the expression of $\sigma(\omega,\q)$ is more complicated. It can be obtained from the analytic propagator  (see e.g. \cite{Blaizot:2001nr})
\beq\label{spectralrepres}
\Delta(\omega,\q)=-\frac{1}{\q^2+\Pi_L(\omega,\q)}+\frac{1}{\q^2}=\int_{q_0} \frac{\sigma(\omega,\q)}{q_0-\omega}
\eeq
with $\Pi_L$ the longitudinal self-energy
\beq
\Pi_L(\omega,\q)=m_D^2 \left(1-\frac{\omega}{2 q}\ln\left( \frac{\omega+q}{\omega-q}  \right)   \right).
\eeq
The imaginary part of $\Pi_L$ (obtained by setting $\omega\to \omega+i\eta$, with $\omega$ real) determines the continuum part of the spectral function at small energy. It is given by
\beq
{\rm Im}\Pi_L(\omega,\q)=\frac{\pi m_D^2 \omega}{2 q}\theta(q-|\omega|).
\eeq
More generally, we have 
\beq
\sigma(\omega,\q)=\frac{2{\rm Im}\Pi_L(\omega,\q)}{ (\q^2+{\rm Re}\Pi_L(\omega,\q))^2+({\rm Im}\Pi_L(\omega,\q))^2}.
\eeq
Note that the temperature enters the spectral function only through the Debye mass $m_D$, and we can set $\sigma(\omega,\q)=m_D^{-2}\,\bar \sigma(\omega/m_D, \q/m_D)$, where $\bar \sigma(\omega/m_D, \q/m_D)$ is a dimensionless function.  On the other hand, the statistical factor that multiplies $\sigma(\omega,\q)$ in Eq.~(\ref{relationsdeltasigma}) depends only on $T$.

\vspace{-0cm}
\begin{figure}[!hbt]
\begin{center}
\includegraphics[width=0.75\textwidth]{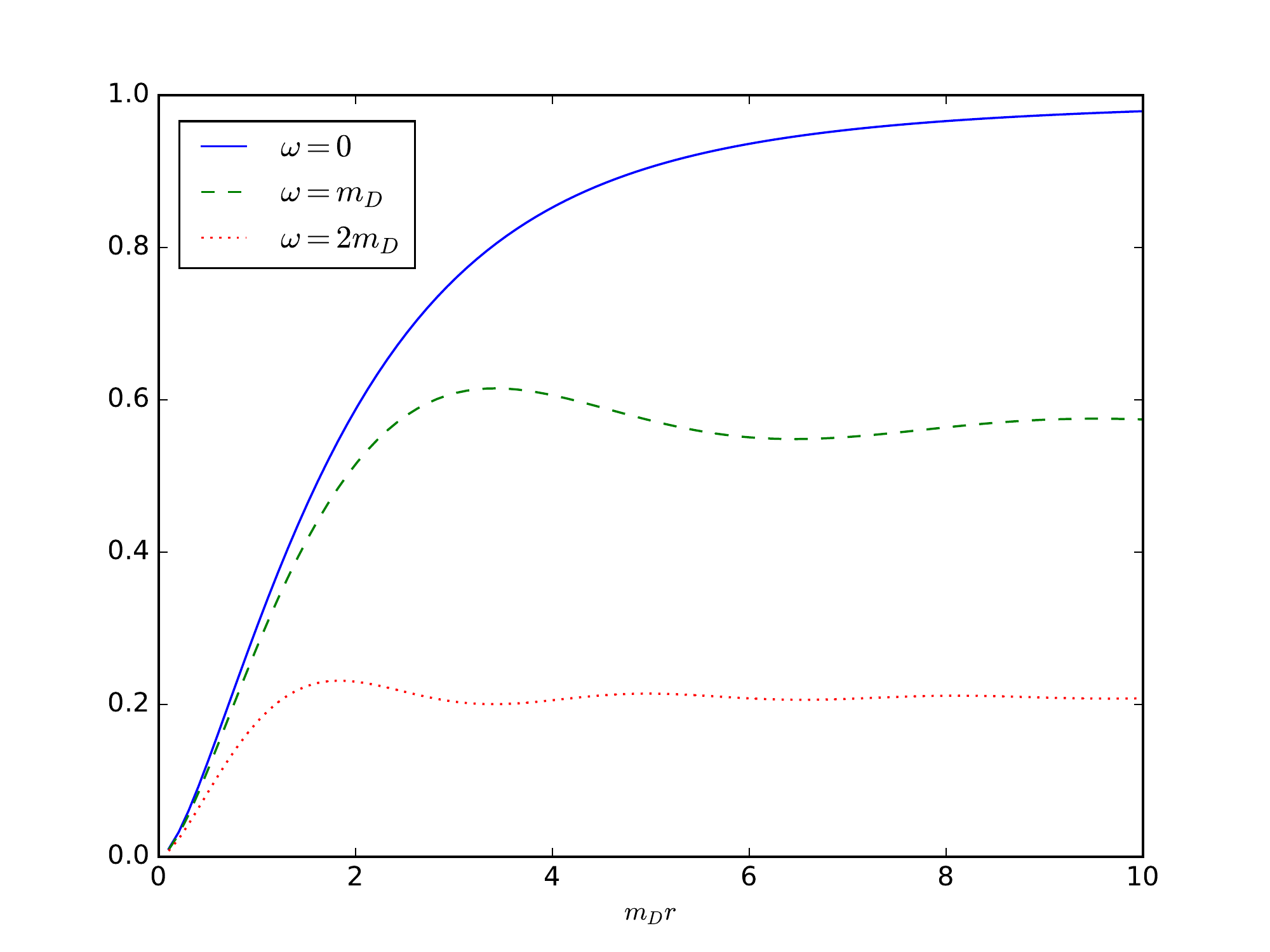}\\
\includegraphics[width=0.75\textwidth]{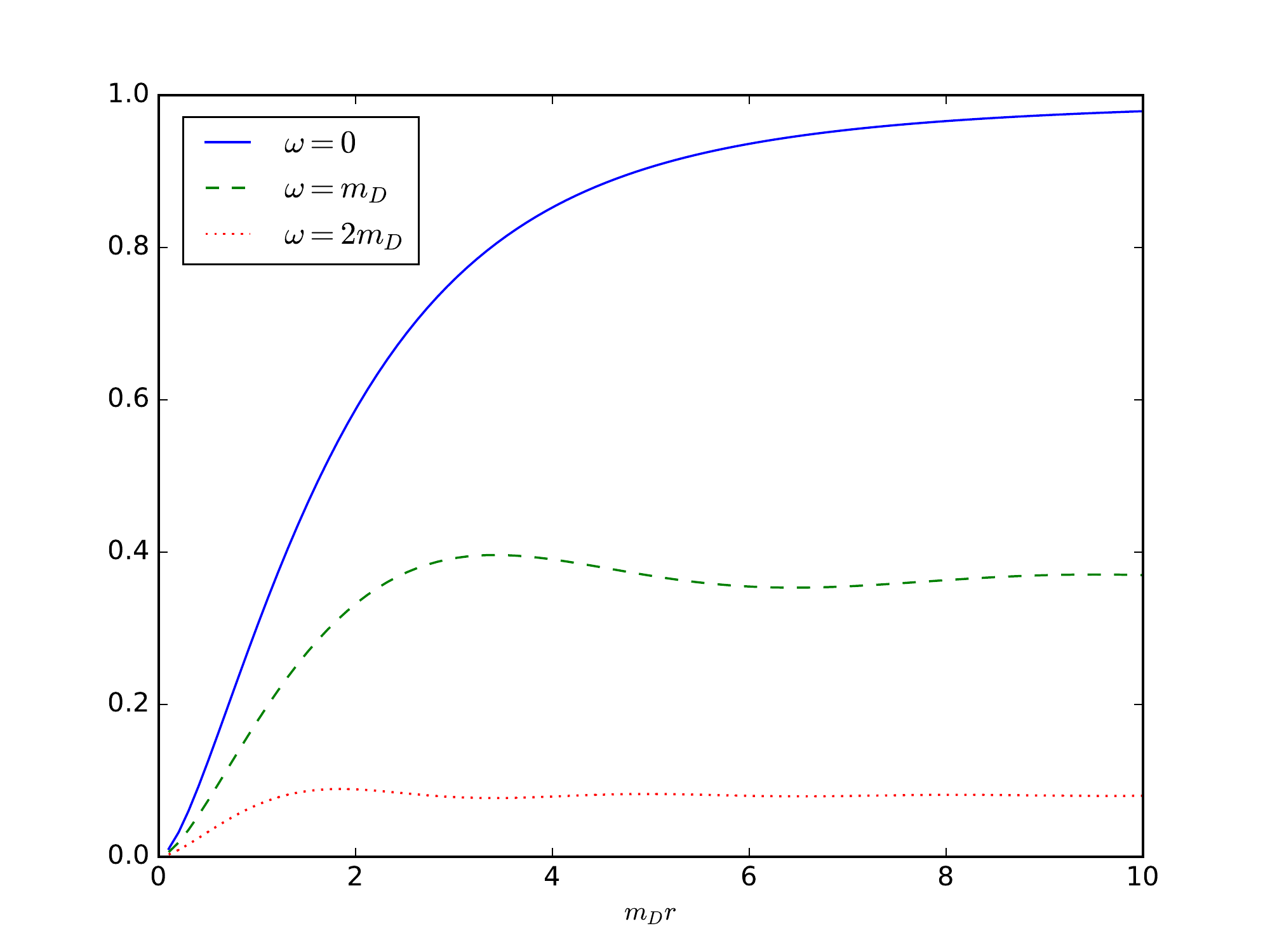}
\vspace{-0cm}
\caption{The function $(4\pi/Tg^2) (W(\omega,r)-W(\omega,0))$ as a function of $rm_D$ for two different temperatures. Top: $T=5m_D$, bottom: $T=m_D$.   } \vspace{-0.25in}
\label{fig:WsE}
\end{center}
\end{figure}

Knowing the spectral function, one can then determine the energy dependent potential (\ref{eq:ImVresum}). The results of this calculation are displayed in Fig.~\ref{fig:WsE} for two values of the temperature, $T=m_D$ and $T=5m_D$. One sees there that the dominant effect of the energy dependence is a sizeable reduction of the imaginary potential, a reduction which gets amplified as the temperature, when it is of the order of the energy, decreases. This reduction arises from the fact that as the energy $\omega$ of the transitions increases the phase space of the  space-like gluons that induce such transitions decreases. The density of such gluons in momentum space is essentially the quantity $\Delta^<(\omega,\q)$ and its decrease  with increasing $\omega$, for a given $q$, results from the combination of two effects:   the statistical factor suppresses transitions with $\omega>T$, and the spectral density vanishes when $\omega>q$. 

To proceed further,  it is convenient to  rewrite Eq.~(\ref{eq:ImVresum}) in terms of dimensionless variables, as follows
\beq
\frac{W(\omega,{\boldsymbol r})-W(\omega,0)}{g^2T}&=&h\left(\frac{\omega}{T}\right)\,\frac{1 }{\bar\omega}\int\frac{\,d^3 \bar q}{(2\pi)^3}\bar \sigma(\bar \omega,\bar {\bf q})\left(1-\rme^{i\bar \q\cdot\bar{\boldsymbol r}}   \right),
\label{eq:ImVresum2}
\eeq
where
\begin{equation}
h(x)\equiv \frac{x}{e^x-1}.
\end{equation}
The curves in Fig.~\ref{fig:WsE} are obtained after integration over $q$, which affects the dependence on $r$ of $W(\omega,r)$ at fixed $\omega$. In particular, at large values of $r m_D$, the last term in the integral (\ref{eq:ImVresum2}) yields a vanishing contribution, which is the origin of the flat behavior observed in Fig.~\ref{fig:WsE}.  Another factor determines the $r$ dependence of the rates: the energy $\omega$ is to be set equal to $\Delta V(r)$ (see Eqs.~(\ref{ratesGammasos})). It turns out that, after this substitution,  the dominant ${\boldsymbol r}$-dependence, in the relevant range, is captured by the function $h$ in Eq.~(\ref{eq:ImVresum2}), that is
\beq
W(\Delta V,{\boldsymbol r})-W(\Delta V,0)&\simeq& h\left(\frac{\Delta V}{T}\right)g(\omega=0,{\bf r})\nn
&\simeq&h\left(\frac{\Delta V}{T}\right) \Gamma_{\rm s}({\boldsymbol r}),
\eeq
where, after reinstating the appropriate color factor $C_F$, we have set
\begin{equation}
g(\omega,{\bf r})\equiv\frac{g^2C_FT}{2\omega}\int_\q \sigma(\omega,\q)\left({\cal S}_{\q\cdot\hat{\boldsymbol r}} \right)^2, 
\end{equation}
and $\Gamma_{\rm s}(r)$ is given explicitly in Eq.~(\ref{Gammarms}).

\begin{figure}
\includegraphics[angle=90,scale=.40]{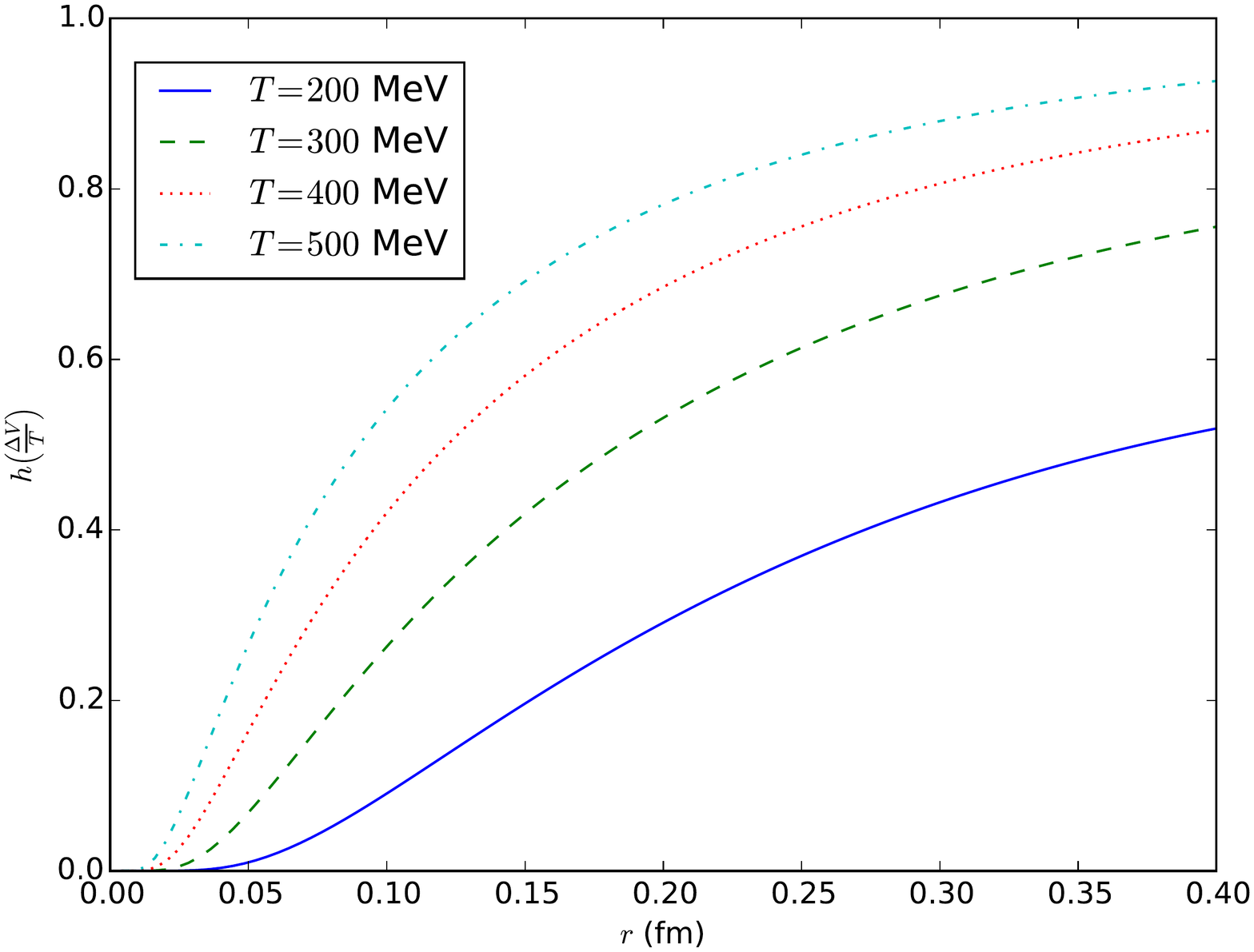}
\caption{The function $h\left(\frac{\Delta V}{T}\right)$ against $r$ for different temperatures}
\label{fig:hvsr}
\end{figure}



The suppression   factor $h\left(\frac{\Delta V({\boldsymbol r})}{T}\right)$ is   plotted as a function of $r$ for different temperatures  in Fig. \ref{fig:hvsr}. Note that $h(x)\to 1$ as $x\to 0$, while, $h(x)\sim \rme^{-x}$ as $x\to \infty$. Thus, the strong suppression at small $r$ originates from the fact that $\Delta V(r)\to\infty $ as $r \to 0$, that is, $h(r)\sim \exp\{-\Delta V(r)/T\}$. This overwhelms the suppression already present in $\Gamma_{\rm s}(r)\sim r^2 \ln(1/r)$, reflecting color transparency, i.e. the suppression of interactions when the size of the color dipole made by the quark and the antiquark in a color singlet vanishes. At large $r$, $\Delta V(r)\to 0$, and $h(r)\sim 1-\Delta V(r)/(2T)$.\\

The setup discussed in this subsection is very close to that used in lattice QCD calculations of the potential, or free energy, of a heavy quark-antiquark pair. In particular,  we  may attempt a comparison with the recent results of Ref.~\cite{Burnier:2015tda}. Since the potential calculated there  is reconstructed from the spectral function, it should  contain, in principle, the energy dependence that we have been discussing. A comparison with the lattice results show that, as is the case with the imaginary potential that we calculate,  the small $r$ dependence is clearly different from the  behavior ($\sim -r^2\log(r)$) expected in the absence of energy dependence: there is a strong suppression at small distances which carries on up to larger radius as the temperature decreases. Unfortunately, a more quantitative comparison between our computation and the lattice simulations is difficult, since the Coulomb approximation that we use is not accurate at large distance, and an additional effect due to the string tension cannot be excluded, as discussed in \cite{Burnier:2015tda}. \\
\begin{figure}
\vspace{-1cm}
\begin{center}
\includegraphics[scale=.4,angle=90]{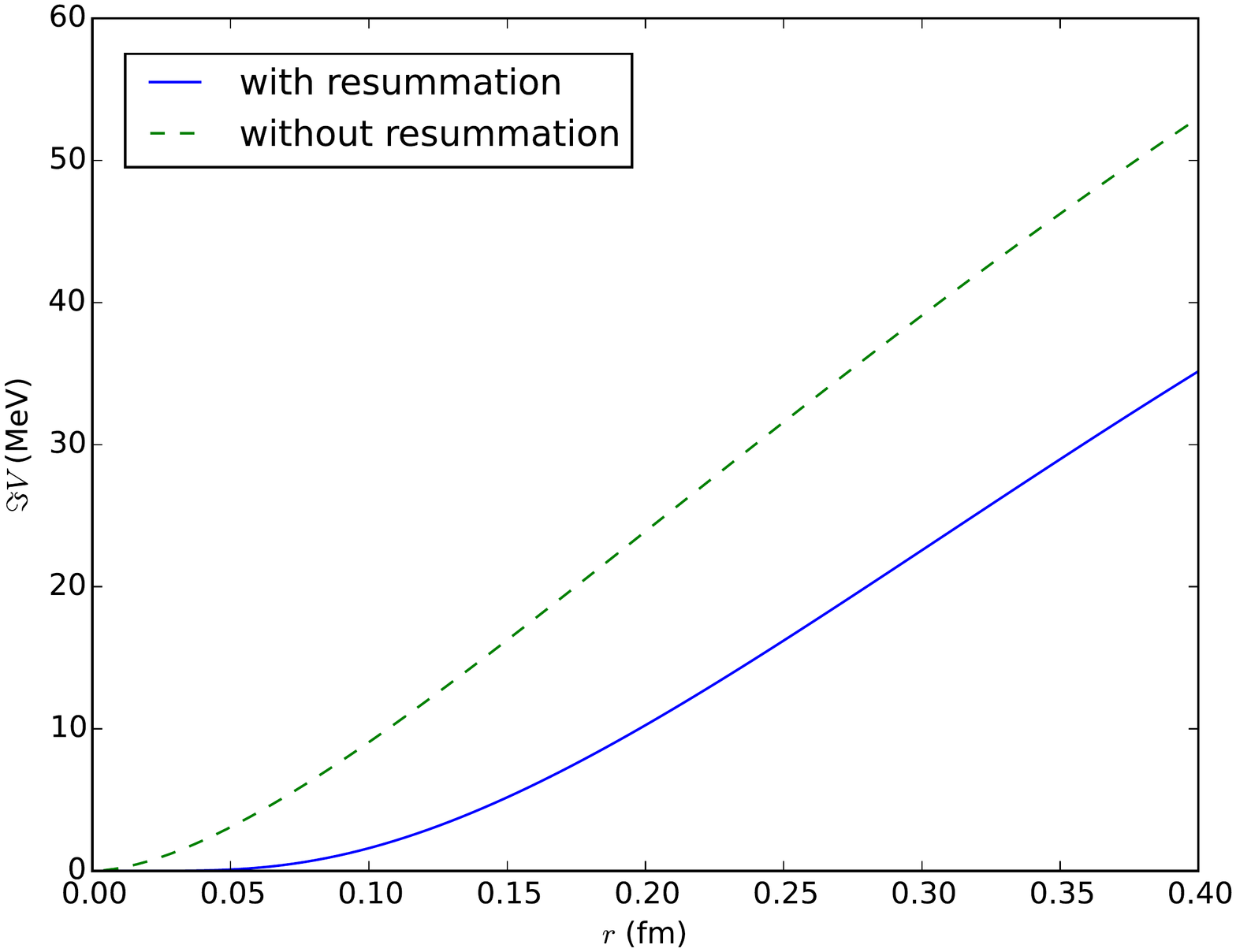}
\caption{Comparison of the imaginary potential with and without energy dependence,  at $T=250\,\textit{MeV}$. }
\label{fig:ImVcom}
\end{center}
\end{figure}

Finally, we return to the estimates of the rate of entropy or free energy variations for which expressions were derived in Sect.\ref{sec:SQED}. The explicit expression of $\Gamma({\boldsymbol r}) $ in the HTL approximation is given in Eq.~(\ref{GammaHTL}). 
Using this approximation we estimate that at $T=250$ MeV,  and using as Bohr radius $a_0=\frac{1}{1200}$ MeV$^{-1}$ the time scale that characterizes the changes in the entropy is around $12\,\rm{fm/c}$. This is of the order  of the  typical total lifetime of the fireball  produced in heavy ion collisions. This suggests that the state of quarkonium will be substantially modified but perhaps not fully thermalized.  Note also that, according to Fig.\ref{fig:ImVcom},  the energy dependence reduces the rate by about a factor 3.

\subsection{A simplified model of quarkonium evolution}\label{sec:quarkoniummodel}

We now move away from the static limit and consider a more ``realistic'' model for the quarkonium. This is based on the following assumptions:
\begin{itemize}
\item We neglect the quark-antiquark interaction in the  octet channel, i.e., we set $V_{\rm o}=0$. This approximation would  be justified in a large $N_c$ limit, and it was used in the original derivation of the gluon-dissociation cross-section \cite{Bhanot:1979vb}. Comparison with later derivations shows that it remains a reasonable approximation for $N_c=3$ \cite{Brambilla:2011sg,Brezinski:2011ju}.
\item This implies in particular that the heavy particles behave as free particles once they are in the octet channel, and octet to octet transitions can be treated in the high temperature limit. In this case, the corresponding equation of motion for the heavy quarks will reduce to a Langevin type equation. 
\item We assume that a single bound state exists in the singlet channel. The survival probablity of this singlet bound state is entirely controlled by its interaction with octet (continuum) states.  In fact, we also ignore continuum singlet states (these represent only about 10\% of the available continuum states). 
\end{itemize}

In summary the model that we consider consists in one bound state, the singlet state, and a continuum of free, octet,  states. We want to see how our equations describe the approach to equilibrium of this particular system. \\

We start with elementary remarks concerning  the system  when it is in thermal equilibrium with the plasma. Recall that we ignore the center of mass motion.  We write the partition function of the relative motion as follows
\beq
Z&=&\rme^{|E_{\rm s}|/T}+\Omega (N_c^2-1)\int_\p \rme^{-\frac{p^2}{MT}}\nn
&=&\rme^{|E_{\rm s}|/T}+\frac{\Omega(N_c^2-1)}{\lambda_T^3},
\eeq
where $|E_{\rm s}|$  is the binding energy ($E_{\rm s}<0$), $\Omega$ is the volume of the plasma, and $\lambda_T=\sqrt{4\pi/MT}$ is the thermal wavelength of the relative motion. At low temperature, the bound state dominates, and $Z\approx\rme^{|E_{\rm s}|/T}$, while at high temperature $Z\approx \Omega(N_c^2-1)/\lambda_T^3$. Clearly, there is a transition temperature $T_c$, of the order of $|E_{\rm s}|$,  corresponding to the situation where these two contributions are of the same order of magnitude, 
\beq
\rme^{|E_{\rm s}|/T_c}=\frac{\Omega(N_c^2-1)}{\lambda_{T_c}^3}.
\eeq
Note that $T_c$ has a (weak, logarithmic) dependence on the volume, and would vanish in an infinite volume. 
Let $p^{\rm s}$ and $p^{\rm o}$ be the probabilities for the system to be in the ground state or in a continuum state, respectively. We have
\beq\label{pspo}
p^{\rm s}_{\rm eq}=\frac{\rme^{|E_{\rm s}|/T}}{Z},\qquad p^{\rm o}_{\rm eq}=\frac{{\Omega(N_c^2-1)}/{\lambda_{T}^3}}{Z}.
\eeq
Clearly, when $T\ll T_c$, $p^{\rm s}\approx 1$, while $p^{\rm o}\approx 1$ when $T\gg T_c$.

The time evolution of the probabilities are given by the simplified system of equations 
\begin{equation}\label{dpsdt}
\frac{\rmd p^{\rm s}}{\rmd t}=g^2C_F\int_\p\left(p_\p^{\rm o}-p^{\rm s}\rme^{-\frac{E_\p^{\rm o}-E^{\rm s}}{T}}\right)\int_\q\Delta^>(\omega_\p^{\rm o}-E^{\rm s},{\q})|\langle {\rm s}|{\cal S}_{\q\cdot\hat{\boldsymbol r}} |{\rm o},\p\rangle|^2\,,
\end{equation}
and
\beq\label{dpodt}
&&\frac{\partial p^{\rm o}_\p}{\partial t}-\gamma{\bf\nabla}({\bf p}p^{\rm o}_\p)-\frac{T\gamma M}{2}\Delta^2 p^{\rm o}_\p=\nn
&&-\frac{g^2}{2N_c}\frac{1}{\Omega}\left(p_\p^{\rm o}-p^{\rm s}\rme^{-\frac{E_\p^{\rm o}-E^{\rm s}}{T}}\right)\int_\q \Delta^>(\omega_\p^{\rm o}-E^{\rm s},{\q})|\langle {\rm s}|{\cal S}_{\q\cdot\hat{\boldsymbol r}} |{\rm o},\p\rangle|^2\,,
\eeq
where $E_\p^{\rm o}=\p^2/M$ is the kinetic energy of the relative motion. We assume that the medium is contained in a cubic box of volume $\Omega$. Computations are made for two different volumes, $\Omega=1\,\rm{fm}^3$ and $\Omega=100\,\rm{fm}^3$ (these values cover the orders of magnitude of the typical volumes of the fireballs produced in a heavy-ion collision). This volume factor affects the numerical results, and it has been made explicit in Eq.~(\ref{dpodt}). Thus,  the plane wave in the equation above is normalized so that  $\langle {\bf r}|{\rm o},\p\rangle=\rme^{i{\bf r}\cdot{\bf p}}$.

The first equation  expresses the change in the bound state population, with a loss term caused by the singlet to octet transitions, while the gain term represents the possible transitions of any of the continuum octet states to the bound singlet. The second equation accounts in addition for the Brownian motion of the particles in the continuum. The specific form of the Langevin terms in the left hand side is taken from Ref.~\cite{Blaizot:2017ypk}.
As a simple consistency check of these equations, one may verify that 
\begin{equation}
\frac{\rmd}{\rmd t}\left(p^{\rm s}+\Omega(N_c^2-1)\int_\p p^{\rm o}_\p\right)=0\,,
\end{equation}
and that the steady state solution is given by Eqs.~(\ref{pspo}). 

We have solved Eqs.~(\ref{dpsdt}, \ref{dpodt}), taking for  $\gamma$ the value used in Ref.~\cite{Blaizot:2015hya},  but adapted  to the bottomonium mass, assuming that $\gamma$ goes as the inverse of the mass, that is,  $\gamma=0.060\,\rm{fm}^{-1}$. Other needed inputs are the binding energy and the wave function of the singlet ground state (that enters the computation of the matrix element $\langle {\rm s}|{\cal S}_{\q\cdot\hat{\boldsymbol r}} |{\rm o},\p\rangle$). We obtain these by solving the Schr\"{o}dinger equation with a screened potential \footnote{This is done with the algorithm of \cite{Lucha:1998xc}, using its python implementation by Hector Martinez (https://github.com/heedmane/schroepy).}
\begin{equation}
V_{\rm s}=-C_F\alpha_s(1/a_0)\frac{e^{-m_D r}}{r}.
\end{equation}
 The results are shown in table \ref{tab:T}.
 \begin{table}[h]
\begin{center}
\begin{tabular}{|c|c|c|c|c|}
\hline
$T(\rm{MeV})$ & $E^{\rm s}(\rm{MeV})$ & $m_D(\rm{MeV})$ & $\alpha_s(2\pi T)$&$\Gamma_{{\rm s}\to {\rm o}} $ (MeV)\\
\hline
$200$ & $-138.36$ & $570.95$ & $0.432$& 6.2 (1.9) \\
$400$ & $-51.57$ & $955.15$ & $0.302$ &39 (13) \\ 
\hline
\end{tabular}
\caption{Table showing different parameters that are used in our simulation at two different temperatures. Note that at $T=0$, i.e., in the vacuum, the binding energy of the singlet state is $E^{\rm s}=-372$ MeV. The last column gives the decay width, $\Gamma_{{\rm s}\to {\rm o}} =a_0^3\int\rmd^3 \p f(pa_0)$, with $f $ given in Eq.~(\ref{fdef}). The values in parenthesis are obtained by using for the evaluation of $f(pa_0)$ the vacuum singlet bound state energy and wave function.}
\label{tab:T}
\end{center}
\end{table}
As can be seen in this table, screening substantially reduces the binding energy. Note that this reduction of the binding energy, together with the corresponding modification of the singlet wave function,  entail a substantial increase of the decay width at a given temperature. This is of course in line with the energy dependence of the rates that we analyzed in the previous section. At $T=400$ MeV, the decay width is of the same magnitude as the binding energy, suggesting that at this temperature the singlet can hardly be considered as a bound state anymore. 
 
\begin{figure}
\includegraphics[scale=0.5]{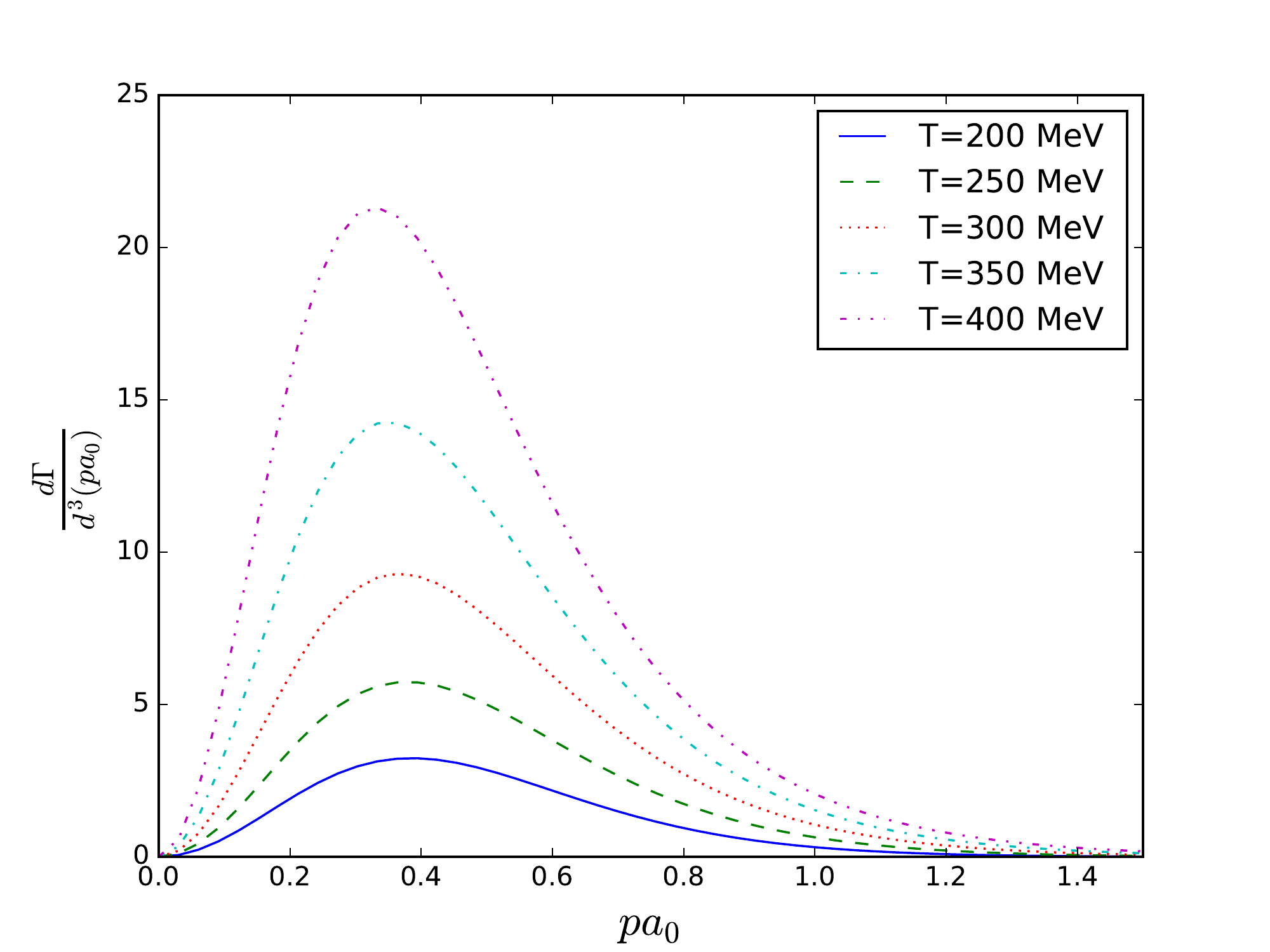}\\
\includegraphics[scale=0.5]{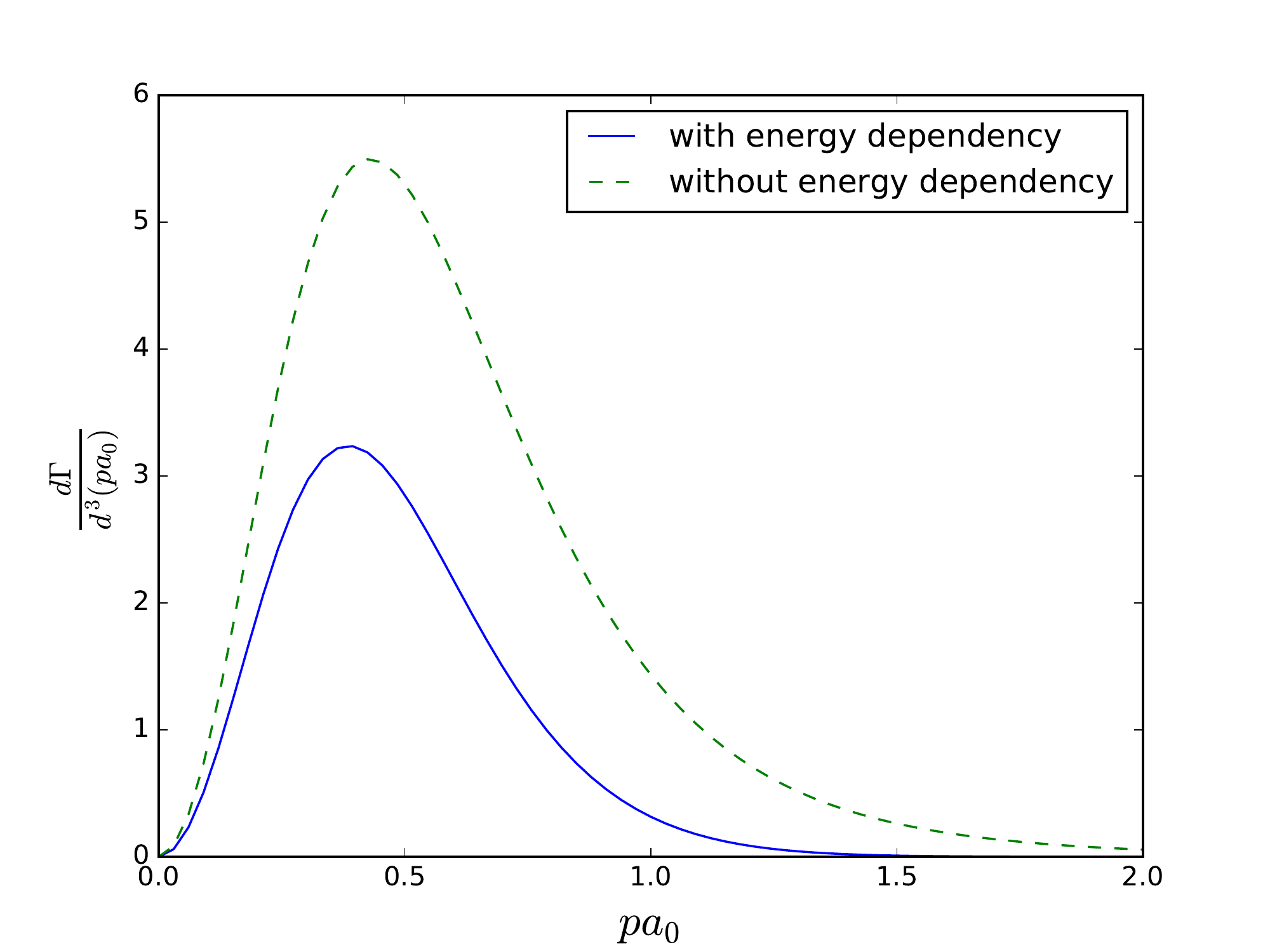}
\caption{Differential decay width  of the singlet bound state into octets as a function of the octet momentum $p$, expressed in units of the inverse of  the Bohr radius $a_0$ of the singlet. The top panel illustrates the dependence on the temperature. The bottom panel illustrates the effect of the energy dependence, at temperature $T=200$ MeV (the curve labelled ``without energy dependence'' corresponds to the function $f(pa_0)$ calculated by substituting $\frac{p^2}{M}-E^s\to 0$).  }
\label{fig:difdecay}
\end{figure}

It is useful to introduce the following  function, proportional to the differential decay rate of a singlet into an octet with momentum $\p$, 
\begin{equation}\label{fdef}
f(pa_0)=\frac{g^2C_F}{(2\pi a_0)^3}\rme^{-\frac{\frac{p^2}{M}-E^{\rm s}}{T}}\int_\q\Delta^>\left(\frac{p^2}{M}-E^{\rm s},{\bf q}\right)|\langle {\rm s}|S_q|{\rm o},\p\rangle|^2.
\end{equation}
This function can be computed numerically once the singlet wave function is known. We use the HTL approximation to evaluate $\Delta^>(\omega,{\bf q})$ (see formulae in the previous subsection). The result of this computation at different temperatures is shown in Fig.~\ref{fig:difdecay}. As one  can see,  the singlet state decays preferentially  into octets whose momentum is of the order of half the inverse of the Bohr radius. We can rewrite the evolution equations of $p^{\rm s}$ and $p^{\rm o}_\p$ in terms of this function $f$:
\begin{equation}
\frac{\rmd p^{\rm s}}{\rmd t}=a_0^3\int\,\rmd^ 3p\left(e^{\frac{\frac{p^2}{M}-E_{1S}^s}{T}}p^{\rm o}_\p-p^{\rm s}\right)f(pa_0)\,,
\label{eq:evops2}
\end{equation}
and
\beq
&&\frac{\rmd p^{\rm o}_\p}{\rmd t}-\gamma{\bf\nabla}(\p\,p^{\rm o}_\p)-\frac{T\gamma M}{2}\Delta^2 p^{\rm o}_\p=\nn
&&\qquad\qquad-\frac{(2\pi a_0)^3}{(N_c^2-1)\Omega}\left(e^{\frac{\frac{p^2}{M}-E_{1S}^s}{T}}p^{\rm o}_\p-p^{\rm s}\right)f(pa_0).
\label{eq:evopo2}
\eeq
 To solve these equations we use the same numerical methods as in Sect.~5.4 of Ref.~\cite{Blaizot:2017ypk}. The most relevant difference as compared to the case treated in \cite{Blaizot:2017ypk} is that, in the present case, the singlet bound state can decay into octets with different momenta. To include this feature in our simulation,  we use a rejection sampling based on the differential decay width to select the momentum.\\

%
%
%

 The value of the survival probability of the singlet state, $p^{\rm s}$,  obtained by solving these equations is given in the following table, for two different temperatures,  two different interaction times, and two different volumes. For comparison the value of the equilibrium probability is indicated in the last column. 
 \begin{center}
\begin{tabular}{|c|c|c|c|c|c|c|}
\hline
  & \multicolumn{3}{|c|}{$\Omega=1\,\rm{fm}^3$} & \multicolumn{3}{|c|}{$\Omega=100\,\rm{fm}^3$} \\
\hline
  & $5\,\rm{fm/c}$ & $100\,\rm{fm/c}$ & eq. & $5\,\rm{fm/c}$ & $100\,\rm{fm/c}$ & eq. \\
\hline
$T=200\,\rm{MeV}$ & $0.86$ & $0.136$ & $0.0814$ & $0.85$ & $0.0438$ & $0.00089$ \\
$T=400\,\rm{MeV}$ & $0.39$ & $0.0515$ & $0.0175$ & $0.36$ & $0.0002$ & $0.00018$\\
\hline
\end{tabular}
\end{center}

We see that at late times $p^{\rm s}$ becomes very small:  the system is then completely dominated by the octets, the more so the larger the volume.  However, on time scales that are typical of the lifetime of the plasma in a heavy-ion collision,  $\sim 5\,\rm{fm/c}$  a significant amount of singlets survive, the survival probability being essentially independent of the volume. 



In order to get a feeling for the role of the Brownian motion of the free quarks, we have repeated the calculations dropping the Langevin terms in the left hand side of Eq.~(\ref{eq:evopo2}), i.e., keeping only the time derivative.  We obtain then the results listed in the table below: 
\begin{center}
\begin{tabular}{|c|c|c|c|c|c|c|}
\hline
  & \multicolumn{3}{|c|}{$\Omega=1\,\rm{fm}^3$} & \multicolumn{3}{|c|}{$\Omega=100\,\rm{fm}^3$} \\
\hline
  & $5\,\rm{fm/c}$ & $100\,\rm{fm/c}$ & eq. & $5\,\rm{fm/c}$ & $100\,\rm{fm/c}$ & eq. \\
\hline
$T=200\,\rm{MeV}$ & $0.85$ & $0.124$ & $0.0814$  & $0.85$ & $0.0442$ & $0.00089$\\
$T=400\,\rm{MeV}$ & $0.40$ & $0.0224$ & $0.0175$ & $0.38$ & $0.001$ & $0.00018$\\
\hline
\end{tabular}
\end{center}
While the  Brownian motion does not affect much $p^{\rm s}$ at small times, it produces a momentum broadening that  tends to slow down the disappearance of the singlet bound state at late times.

We have also repeated the calculation, ignoring the energy dependence of the imaginary potential. The results are given in the table below, for the case $T=200$ MeV. 
\begin{center}
\begin{tabular}{|c|c|c|c|c|}
\hline
  & \multicolumn{2}{|c|}{$\Omega=1\,\rm{fm}^3$} & \multicolumn{2}{|c|}{$\Omega=100\,\rm{fm}^3$} \\
\hline
  & $5\,\rm{fm/c}$ & $100\,\rm{fm/c}$ & $5\,\rm{fm/c}$ & $100\,\rm{fm/c}$  \\
\hline
$T=200\,\rm{MeV}$ & $0.5631$ & $0.0093$ & $0.5596$ & $0.001$ \\
\hline
\end{tabular}
\end{center}
As can be seen from this table, the survival probability is much reduced, and it eventually vanishes at large time: the absence of an energetic penalty for the transition to an octet state allows for a rapid occupation of the large continuum phase space. This provides another indication of the importance of the  energy dependence of the rates.

As a final remark, we have evaluated the free energy, internal energy and the entropy in equilibrium. These are shown in the following table 
\begin{center}
\begin{tabular}{|c|c|c|c|c|c|c|}
\hline
  & \multicolumn{3}{|c|}{$\Omega=1\,\rm{fm}^3$} & \multicolumn{3}{|c|}{$\Omega=100\,\rm{fm}^3$} \\
\hline
  & $F\,\rm{MeV}$ & $E\,\rm{MeV}$ & $TS\,\rm{MeV}$ & $F\,\rm{MeV}$ & $E\,\rm{MeV}$ & $TS\,\rm{MeV}$  \\
\hline
$T=200\,\rm{MeV}$ & $-640.07$ & $264.32$ & $904.39$ & $-1544.30$ & $299.61$ & $1843.91$ \\
$T=400\,\rm{MeV}$ & $-1669.14$ & $588.58$ & $2257.72$ & $-3504.21$ & $599.88$ & $4104.09$ \\
\hline
\end{tabular}
\end{center}
In all cases, the free energy is dominated by the entropy, the more so the larger $T$ and/or $\Omega$. Note that in the large volume $\Omega$ limit $<E>$ goes to a constant, while $F\sim -TS$ goes to infinity but just as -$T\log\Omega$.

\section{Conclusions} 

The equations for the reduced density matrix that we have derived in this paper describe the evolution of a quarkonium towards thermal equilibrium in both regimes of high and moderate temperatures. The high temperature regime is that where the binding energies can be ignored. Then the dynamics is well described by a Lindblad equation and the approach to equilibrium is controlled by the increase of the von Neumann entropy. In this regime, binding forces can be treated perturbatively, and the effects of the collisions accounted for by a static imaginary potential. In the regime of moderate temperature, binding energies cannot be ignored, and it is convenient to use as a reference basis, that of the eigenstates of the effective heavy quark hamiltonian. The equation for the density matrix then leads to rate equations describing transitions between these eigenstates, and the approach to equilibrium is accompanied by the decrease of a free energy. The dynamics of continuum states  remains dominated by Brownian motion and is described by a Langevin equation. In this regime of moderate temperature, the effect of collisions is still captured by an imaginary potential, which enters the determination of the collision rates. An important feature of the imaginary potential is that it depends on the energy: this is because as the energy of a transition increases, the phase space of the space-like gluons that cause this transition decreases. This energy dependence is found to be numerically important and should be taken into account in phenomenological studies. As we have emphasized, this effect is expected to be much more important for QCD than it would be for QED. This is because the absorption of a gluon by a quark-antiquark pair changes the color state of the pair, and turns an attractive force into a repulsive one, or vice versa. 

The last section of the paper presented numerical studies illustrating the main concepts discussed in the earlier sections. The first example is that of a pair of infinitely massive quark and antiquark. This is close to the typical set up used in lattice gauge calculations, and some comparison with recent lattice results has been attempted. It would certainly be worthwhile to extend such comparison and see whether the strong suppression arising from  the energy dependence of the imaginary part of the potential at small separation can be reproduced by lattice calculations. The second example treats a simplified model of a quarkonium with a single bound state in a singlet state, and continuum octet states. Although this is an oversimplification of the realistic situation, many interesting features emerge from this study, that could be relevant in phenomenological studies. This example illustrates in particular the interplay between screening and collisions, and the importance of treating both on the same footing, as we do in this paper.\footnote{The effect of the change in the binding energies on the collision rates is taken into account in some recent phenomenological analysis, such as that based on the in-medium T-matrix \cite{Du:2017qkv}.}

In this paper we have   focussed on a simple question, how a quarkonium approaches equilibrium when it is in contact with a static quark-gluon plasma in thermal equilibrium at temperature $T$. Although we have examined only simplified models, the equations that we have derived allow in principle  for a quantitative answer to this question. They should provide a consistent starting point for more elaborate phenomenological work. The formalism developed in this paper should be well suited to the study of the bottomonium, presumably to be found in the moderate temperature regime in heavy ion collisions. The case of charmonium is more intricate, and presumably calls for a mix of techniques, in particular if one wishes to address the issue of recombination. Then the approximations developed in \cite{Blaizot:2017ypk} may be useful.

 \section*{Acknowledgements}
 This work has been supported in part by the European Research Council under the
Advanced Investigator Grant ERC-AD-267258. The work of M.A.E. was supported by the Academy of Finland, project 303756.

\appendix
\section{Elimination of the center of mass coordinates \label{sec:cm}}
In this appendix, we perform the Fourier transform of our main equation, and eliminate the center of mass coordinate. 

In order to proceed with the Fourier transform,   we note that the correlator $\Delta_-^<({X},{X}')$ defined in Eq.~(\ref{deltabigminus})  depends on the difference of times, $\tau=t-t'$, and a priori on 4 coordinates. Because of translation invariance, it is in fact function of only three differences of coordinates. To make this more explicit, we consider  $\Delta_-^<({X},{X}')$ as an operator in the two particle space, with matrix elements 
\beq
&&\bra{{\boldsymbol r}_1,\bar{\boldsymbol r}_1}\Delta_-^<(\tau,\X,\X')\ket{{\boldsymbol r}_2,\bar {\boldsymbol r}_2}\nn
&&\qquad\qquad=\Delta^<(\Y+\y/2)+\Delta^<(\Y-\y/2)-\Delta^<(\Y+{\boldsymbol r})-\Delta^<(\Y-{\boldsymbol r})\nn
&&\qquad\qquad\equiv\Delta_-(\tau;\Y,\y,{\boldsymbol r}),
\eeq
where 
\beq
&&{\boldsymbol R}_i\equiv\frac{{\boldsymbol r}_i+\bar{\boldsymbol r}_i}{2},\qquad \s_i\equiv{\boldsymbol r}_i-\bar{\boldsymbol r}_i, \qquad (i=1,2)\nn
&&\Y\equiv{\boldsymbol R}_1-{\boldsymbol R}_2,\qquad \y\equiv\s_1-\s_2,\qquad {\boldsymbol r}\equiv\frac{\s_1+\s_2}{2}.
\eeq
The coordinates $\Y$ and $\y$ play an important role in the semi-classical approximation (see \cite{Blaizot:2017ypk}). Thus, in the large mass limit, ${\boldsymbol r}_1\approx {\boldsymbol r}_2$ and $\bar {\boldsymbol r}_1\approx \bar{\boldsymbol r}_2$,  $\Y\to 0 $, $\y\to 0$ and ${\boldsymbol r}$ becomes equal to the relative coordinate. It is convenient to express the Fourier transforms in terms of these variables. 
We have,  for instance (with the shorthand notation $\int_\q=\int\frac{\rmd ^3\q}{(2\pi)^3}$)
\beq
\Delta^<(\tau,{\boldsymbol r}_1-{\boldsymbol r}_2)&=&\int_\q\rme^{i\q\cdot ({\boldsymbol r}_1-{\boldsymbol r}_2)} \Delta^<(\tau,\q)\nn
&=&\int_\q\rme^{i\q\cdot \Y} \rme^{i\q\cdot \y/2} \Delta^<(\tau,\q),
\eeq
and for $\Delta_-^<(\tau;\Y,\y,{\boldsymbol r})$,
\beq
\Delta_-(\tau;\Y,\y,{\boldsymbol r})=2\int_\q\rme^{i\q\cdot \Y} \left[ \cos(\q\cdot \y/2)-\cos(\q\cdot{\boldsymbol r})   \right]\Delta(\tau,\q).
\eeq
The variable $\q$ has the meaning of the momentum of the exchange gluon. 
We can also take a Fourier transform with respect to the time variable ($q=(q_0,\q)$)
\beq
\Delta(\tau,\q)=\int_{q_0} \rme^{-i q_0\tau} \Delta(q),
\eeq
and write (with now $\int_q=\int\frac{\rmd ^4 q}{(2\pi)^4}$)
\beq
\Delta_-(\tau;\Y,\y,{\boldsymbol r})=2\int_q \Delta_-(q,\y, {\boldsymbol r}) \rme^{-iq_0\tau}\rme^{i\q\cdot \Y}.
\eeq
We shall sometimes write, with a slight abuse of notation,  $\Delta_-(q,\s_1, \s_2)$ in place of $\Delta_-(q,\y, {\boldsymbol r})$. To summarize, we  can write $\bra{{\boldsymbol r}_1,\bar{\boldsymbol r}_1}\Delta_\pm^<(\tau,\X,\X')\ket{{\boldsymbol r}_2,\bar {\boldsymbol r}_2}$ as
\begin{equation}
\Delta_\pm^<({\boldsymbol R}_1,{\boldsymbol R}_2,\s_1,\s_2;\tau)=2\int\frac{\rmd^4q}{(2\pi)^4}\Delta_\pm^<(q,\s_1,\s_2)\,\rme^{-iq_0\tau+i\q\cdot({\boldsymbol R}_1-{\boldsymbol R}_2)}\,,
\end{equation}
where $q=(q_0,{\bf q})$  has the interpretation of the exchanged gluon four-momentum, and
\begin{equation}
\Delta_\pm^<(q,\s_1,\s_2)=\Delta^<(q)\left[\cos\left(\frac{{\bf q}(\s_1-\s_2)}{2}\right)\pm\cos\left(\frac{{\bf q}(\s_1+\s_2)}{2}\right)\right].
\end{equation}
Similar relations hold for  $\Delta_\pm^>({\boldsymbol R}_1,{\boldsymbol R}_2,\s_1,\s_2;\tau)$.\\

These relations allow us to perform the partial trace over center of mass coordinate, or equivalently over the center of mass momentum. 
We illustrate the procedure with the first term of  Eq.~(\ref{DsDoME}), and more precisely the first contribution to Eq.~(\ref{eq:evosin1}).
Consider then the matrix element
\beq\label{melR1R2}
\bra{{\boldsymbol R}_1,\s_1}{\cal P}_\X U_{\rm o}(\tau){\cal P}_{\X'} (\tau)D_{\rm s}(t-\tau)U_{\rm s}^\dagger(\tau)\ket{{\boldsymbol R}_2,\s_2}\Delta^>_-(X,X')
\eeq
between  localized states of the heavy quark antiquark pair (in a color singlet state). Taking advantage of the fact that the projectors are diagonal in coordinate space, we can rewrite this as (omitting to indicate time variables to simplify the writing)
\beq
\int_{\X_2',\bar\X_2}\bra{\X_1} U_{\rm o}\ket{\X_2'} \bra{\X_2'} D_{\rm s}\ket{\bar \X_2}\bra{\bar \X_2}U_{\rm s}^\dagger\ket{\X_2}\Delta^>_-(\tau,\X_1,\X_2')\nn
\eeq
where $\X_i=({\boldsymbol R}_i,\s_i)$. We note then that the evolution operators factorize into a center of mass contribution which depends only on the kinetic energy of the center of mass, and a part related to the relative motion that involves the potentials $V_{\rm s}$ or $V_{\rm o}$. We set $U_{\rm s}=U_{\rm cm}\tilde U_{\rm s}$ and similarly for $U_{\rm o}$.  We have
\beq
\bra{\X_1} U_{\rm o}(\tau)\ket{\X_2'} &=&\bra{{\boldsymbol R}_1,\s_1} U_{\rm cm}\tilde U_{\rm o}\ket{{\boldsymbol R}_2',\s_2'} \nn
&=&\bra{{\boldsymbol R}_1} U_{\rm cm}\ket{{\boldsymbol R}_2'}\bra{\s_1} \tilde U_{\rm o}\ket{\s_2'},
\eeq
where
\beq
\bra{{\boldsymbol R}_1} U_{\rm cm}\ket{{\boldsymbol R}_2'}=\int_{\P_1'}\rme^{-i\tau \frac{\P_1'^2}{4M}}\rme^{i\P_1'\cdot({\boldsymbol R}_1-{\boldsymbol R}_2')},
\eeq
and $\tilde U_{\rm o}$ acts on the relative coordinates, in the octet channel. A simple calculation then yields for the matrix element of Eq.~(\ref{melR1R2}) between center of mass momentum states $P_1$ and $P_2$, 
\beq
\int_{\s_2',\s_2''}\bra{\s_1} \tilde U_{\rm o}\ket{\s_2'} \bra{\P_1,\s_2'}D_{\rm s}\ket{\P_2,\s_2''}\bra{\s_2''} \tilde U_{\rm s}^\dagger\ket{\s_2}\Delta^>_-(\tau;\q,\s_1,\s_2') \rme^{-i\tau \frac{(\P_1-\q)^2}{4M}}\rme^{i\tau \frac{\P_2^2}{4M}}, \nn
\eeq
which can also be written as an operator equation in the space of relative coordinates
\beq
\int_{\s'}{\cal P}_{\s}  \tilde U_{\rm o}(\tau) {\cal P}_{\s'} \bra{\P_1}D_{\rm s}(t-\tau)\ket{\P_2} \tilde U_{\rm s}^\dagger (\tau)\Delta^>_-(\tau;\q,\s,\s')\, \rme^{-i\tau \frac{(\P_1-\q)^2-P_2^2}{4M}}.
\eeq
At this point we recall that 
\beq
\Delta^>_-(q,\s,\s')=2\sin\frac{\q\cdot\s}{2}\,\sin\frac{\q\cdot\s'}{2}\,\Delta^>(q), 
\eeq
so that the first contribution to ${\cal L}^{\rm ss}(\tau) D_{\rm s}(t-\tau)$ in Eq. (\ref{eq:evosin1}) finally reads
\beq\label{eq:evosin1A}
&&-g^2C_F \int_{0}^{t-t_0} \rmd \tau \int_q \rme^{-iq_0\tau}\Delta^>(q)\nn
&&\qquad\times{\cal S}_{\q\cdot\hat\s} \tilde U_{\rm o}(\tau){\cal S}_{\q\cdot\hat\s}  \tilde U_{\rm s}^\dagger(\tau)\bra{\P_1}D_{\rm s}(t)\ket{\P_2}\, \rme^{-i\tau \frac{(\P_1-\q)^2-P_2^2}{4M}},
\eeq
where
\beq
{\cal S}_{\q\cdot \s}\equiv 2\sin(\q\cdot \hat\s/2),
\eeq
and $\hat\s$ is the relative coordinate operator. 

At this point it is (almost) trivial to trace out the center of mass degrees of freedom. This amounts to  set $\P_1=\P_2$ and to integrate over $\P_1$.
Note that the commutator in Eq.~(\ref{DsDoME}) yields 
\beq
&&\bra{\P_1,\s_1}[H_Q,D_{\rm s}]\ket{\P_2,\s_2}\nn
&&\quad\qquad =\frac{\P_1^2-\P_2^2}{4M} \bra{\P_1,\s_1}D_{\rm s}\ket{\P_2,\s_2}+\bra{\P_1,\s_1}[H_{\rm s},D_{\rm s}]\ket{\P_2,\s_2}.
\eeq
The first term will not contribute when taking the trace (with $\P_1=\P_2$). As for the second term, it yields
$\bra{\s_1}[H_{\rm s},\tilde D_{\rm s}]\ket{\s_2}$.

To proceed further we need to analyze the phase factor in Eq.~(\ref{eq:evosin1A}).  This is the  product of $\tau$ by the recoil energy
\beq
\Delta E_{\rm recoil}=\frac{(\P-\q)^2}{4M}-\frac{\P^2}{4M}=\frac{1}{4M}\left( \q^2-2\P\cdot\q   \right),
\eeq
where we have set $P_1=P_2=P$. The quantity $\Delta E_{\rm recoil}$ is the recoil energy of the heavy quark system, after absorption or emission  of a gluon with momentum $\q$. The range of the  $\tau$-integration in Eq.~(\ref{eq:evosin1A}) is limited by the propagator $\Delta^>(\tau,\q)$ to be of the order of the inverse Debye mass $m_D\lesssim T$. On the other hand, the collisions of the heavy particles with the light constituents of the plasma involve the exchange of soft gluons, with $|\q|\lesssim m_D$. It follows that typically, $q\tau\sim 1$, and the recoil energy is a small fraction of the Debye mass, $\tau q^2/M\lesssim (m_D/M)$. A similar estimate holds for the term $\tau\P\cdot \q/M\lesssim (T/M)$, where we have assumed that $P\lesssim T$ (we consider pairs that are initially at rest. If the center of mass momentum is high then we need to consider ``hot wind'' effects, which is beyond the  scope of this paper \cite{Chu:1988wh,Escobedo:2013tca}). Since we assume that both $m_D\ll M$ and $T\ll M$, one can safely ignore the phase factor. 

A similar reasoning can be made for all the contributions to the main equations. We then obtain the equations that are listed in the main text. 

\section{Multiple-scale analysis}\label{secular}

In this appendix, we discuss the solution of Eq.~(\ref{Liouville1}) within perturbation theory, paying particular attention to the secular terms. 
We first   rewrite Eq.~(\ref{eq:evolt3}) as follows
\begin{equation}
\frac{\rmd{\cal D}}{dt}+i[H,{\cal D}]=\epsilon\mathcal{F}[\cal D]\,,
\label{eq:evolt4}
\end{equation}
where ${\cal F}$ is a linear functional of ${\cal D}$ and $\epsilon$  a small parameter.  We regard the right hand side as a perturbation and attempt to solve Eq.~(\ref{eq:evolt4}) as an expansion in powers of $\epsilon$. That is, we write
\begin{equation}
{\cal D}(t)={\cal D}_0(t)+\epsilon{\cal D}_1(t)+\cdots
\end{equation}
and obtain
\begin{equation}
{\cal D}_0(t)=e^{-iHt}{\cal D}_0(0)e^{iHt} \,,
\end{equation}
and
\begin{equation}
{\cal D}_1(t)=\int_0^t\,dt'e^{-iH(t-t')}\mathcal{F}[{\cal D}_0(t')]e^{iH(t-t')}\,.
\end{equation}
The difficulty with this naive expansion is that the condition $\epsilon{\cal D}_1\ll {\cal D}_0$ is not always satisfied. In particular, this condition is violated  at late times  if the  following quantity
\begin{equation}
\mathcal{F}_S[{\cal D}_0(t)]=\lim_{T\to\infty}\frac{1}{T}\int_0^T\,dt'e^{-iH(t-t')}\mathcal{F}[{\cal D}_0(t')]e^{iH(t-t')},
\end{equation}
is not equal to zero. Let us then set
\begin{equation}
\mathcal{F}[{\cal D}_0(t)]=\mathcal{F}_S[{\cal D}_0(t)]+\delta\mathcal{F}[{\cal D}_0(t)]
\end{equation}
so that 
\begin{equation}
{\cal D}_1(t)=te^{-iHt}\mathcal{F}_S[{\cal D}_0(0)]e^{iHt}+\int_0^t\,dt'e^{-iH(t-t')}\delta\mathcal{F}[{\cal D}_0(t')]e^{iH(t-t')}\,.
\end{equation}
This expression makes explicit the secular term, growing linearly with time, at the origin of the breakdown of naive perturbation theory. 

 The problem can be handled by multiple-scale analysis (see e.g. chapter 11 of \cite{Bender}). One introduces a ``slow'' time $\tau=\epsilon t$, and consider $D$ as a function of $t$ and $\tau $, treated (artificially) as independent variables. We have, as before,  
\begin{equation}
{\cal D}(t,\tau)={\cal D}_0(t,\tau)+\epsilon{\cal D}_1(t,\tau)+\cdots
\end{equation}
The leading order equation reads
\begin{equation}
\frac{\partial{\cal D}_0}{\partial t}+i[H,{\cal D}_0]=0\,,
\end{equation}
so that
\begin{equation}
{\cal D}_0(t,\tau)=e^{-iHt}{\cal D}_0(0,\tau)e^{iHt} \,.
\label{eq:D0t}
\end{equation}
The next to leading order equation involves the derivative of $ {\cal D}_0(t,\tau)$ with respect to $\tau$, viz.
\begin{equation}
\epsilon\frac{\partial{\cal D}_0}{\partial \tau}+\epsilon\frac{\partial{\cal D}_1}{\partial t}+i\epsilon[H,{\cal D}_1]=\epsilon\mathcal{F}[{\cal D}_0(t,\tau)].
\end{equation}
We can now use $\frac{\partial{\cal D}_0}{\partial \tau}$ in order to cancel the secular contribution of $\mathcal{F}[{\cal D}_0(t,\tau)]$, that is, we demand that the following equation 
\begin{equation}
\frac{d{\cal D}_0}{d\tau}=\mathcal{F}_S[{\cal D}_0(t,\tau)]
\label{eq:evolt5}
\end{equation}
be satisfied. This fixes the $\tau $ dependence of ${\cal D}_0(t,\tau)$. 
Then we can solve for ${\cal D}_1$,
\begin{equation}
{\cal D}_1(t,\tau)=\int_0^t\,dt'e^{-iH(t-t')}\delta\mathcal{F}[{\cal D}_0(t',\tau)]e^{iH(t-t')}.
\end{equation}
By construction, ${\cal D}_1(t,\tau)$ non longer contains a secular term, and can be considered a genuine perturbative quantity.  
A similar result could have been obtained by applying   renormalization group techniques, as discussed recently in Ref.~\cite{PhysRevE.96.042113}. 

The separation of the secular term requires the solution of Eq.~(\ref{eq:evolt5}). By projecting this equation on the eigenvectors of the operator $H$, assuming that all energy levels are discrete, we obtain
\beq
&&\langle n|\mathcal{F}_S[{\cal D}_0(t,\tau)]|m\rangle=\nonumber\\
&&\lim_{T\to\infty}\frac{1}{T}\int_0^T\,dt'\sum_{n'm'}e^{-it(E_n-E_m)}e^{it'(E_n-E_{n'}-E_m+E_{m'})}\mathcal{F}[{\cal P}_{n'}{\cal D}_0(0,\tau){\cal P}_{m'}].\nn
\eeq
This is non-zero only if $E_n-E_m=E_{n'}-E_{m'}$. Thus  the evolution described by Eq.~(\ref{eq:evolt5}) only connects pairs of states whose energy differences $E_{mn}=E_m-E_n$ are identical. It follows in particular that the evolution of the populations of the various quarkonium states, i.e., of the diagonal elements, for which $E_{mn}=0$, decouples from that of the non-diagonal ones (at leading order  and assuming absence of degenerate states). One can also evaluate similarly the matrix elements of ${\cal D}_1(t,\tau)$. One gets
\beq
\langle n|{\cal D}_1(t,\tau)|m\rangle=i\sum_{n'm'}\frac{\rme^{-itE_{nm}}-\rme^{-it'E_{n'm'}}}{E_{n'm'}-E_{nm}} \delta{\cal F}[\langle n'|D_0(0,\tau)|m'\rangle],
\eeq
where, by construction, $E_{n'm'}\ne E_{nm}$. Assuming that the particles are confined in a volume $\Omega\sim L^3$, the lowest values of the energy denominators are of order $L^{-1}$. Thus, if $\Gamma$ denotes the leading order decay rate, ${\cal D}_1$ will remain a small perturbative correction as long as $\Gamma L\ll 1$. In the example treated in Sect.~\ref{sec:quarkoniummodel}, this condition is well satisfied for $\Omega=1\; {\rm fm}^3$, but only marginally for $\Omega=100\; {\rm fm}^3$.

\bibliography{./freedraft}
\biboptions{sort&compress}
\bibliographystyle{elsarticle-num.bst}
\bibliographystyle{plain}
\end{document}